\documentstyle[epsfig,12pt]{article}

\begin{document}

%%-move to normal A4-%%
\hoffset = -1truecm \voffset = -2truecm \baselineskip = 10 mm

\title{\bf The chaotic effects in a nonlinear QCD evolution equation}

\author{
{Wei Zhu}\footnote{Corresponding author, weizhu@mail.ecnu.edu.cn}, {Zhenqi Shen} and {Jianhong Ruan}\\\\
\normalsize Department of Physics, East China Normal University,
Shanghai 200241, P.R. China}

\date{}

\newpage

\maketitle

\vskip 3truecm

\begin{abstract}

     The corrections of gluon fusion to the DGLAP and BFKL equations
are discussed in a united partonic framework. The resulting
nonlinear evolution equations are the well-known GLR-MQ-ZRS
equation and a new evolution equation.  Using the available
saturation models as input, we find that the new evolution
equation has
the chaos solution with positive Lyaponov exponents
in the perturbative range. We predict a new kind of shadowing
caused by chaos, which blocks the QCD evolution in a critical
small $x$ range. The blocking effect in the evolution equation may
explain the Abelian gluon assumption and even influence our
expectations to the projected Large Hadron Electron Collider
(LHeC), Very Large Hadron Collider (VLHC) and the upgrade (CppC) in a circular $e^+e^-$ collider (SppC).

\end{abstract}

{\bf keywords}:  QCD evolution equation; Chaos; Saturation; Blocking
effect; LHeC; VLHC

{\bf PACS numbers}: 12.38.-t; 14.70.Dj; 05.45.-a

\newpage
\begin{center}
\section{Introduction}
\end{center}

    The QCD evolution equation is an important part in the study of high
energy physics. The linear DGLAP
(Dokshitzer-Gribov-Lipatov-Altarelli-Parisi) equation [1] and BFKL
(Balitsky-Fadin-Kuraev-Lipatov) equation [2] are no longer reliable
at ultra higher energy since the corrections of parton
recombination. A series of nonlinear evolution equations, for
example, the GLR-MQ-ZRS (Gribov-Levin-Ryskin, Mueller-Qiu,
Zhu-Ruan-Shen) equation [3,4] and BK (Balitsky-Kovchegov) equation
[5] were proposed, in which the corrections of parton recombination
are considered.

    As we know, the nonlinear iteration equations may have a characteristic solution--chaos,
which has been observed in many natural phenomena [6]. A following
question is: do the nonlinear QCD evolution equations have chaotic
solution?  Several years ago we have reported chaos in a new
evolution equation [7], which describes the corrections of the gluon
recombination to the BFKL equation at the leading logarithmic
$LL(1/x)$ approximation. The purpose of this work is to detail this
discovery after a long consideration.

    We begin from the proposal of the new evolution equation. Fig. 1
is a schematic program, which shows that the correlations among
initial gluons modify the evolution equations step by step. The
elementary amplitude Fig. 1a together with its conjugate amplitude
constructs the DGLAP equation for gluon. The correlations among the
initial partons are neglected in the DGLAP equation. This assumption
is invalid in the higher density region of partons, where the parton
wave functions begin to spatially overlap. Therefore, the
corrections of the correlations among initial gluons to the
elementary DGLAP amplitude at small $x$ should be considered. To
this end, we add the possible initial gluon lines on Fig. 1a step by
step. The resulting three sets of amplitudes are listed in Fig.
1b-1d. It is interesting that these amplitudes produce the BFKL,
GLR-MQ-ZRS equations and a new evolution equation.

\vskip -1.0 truecm \hbox{
\centerline{\epsfig{file=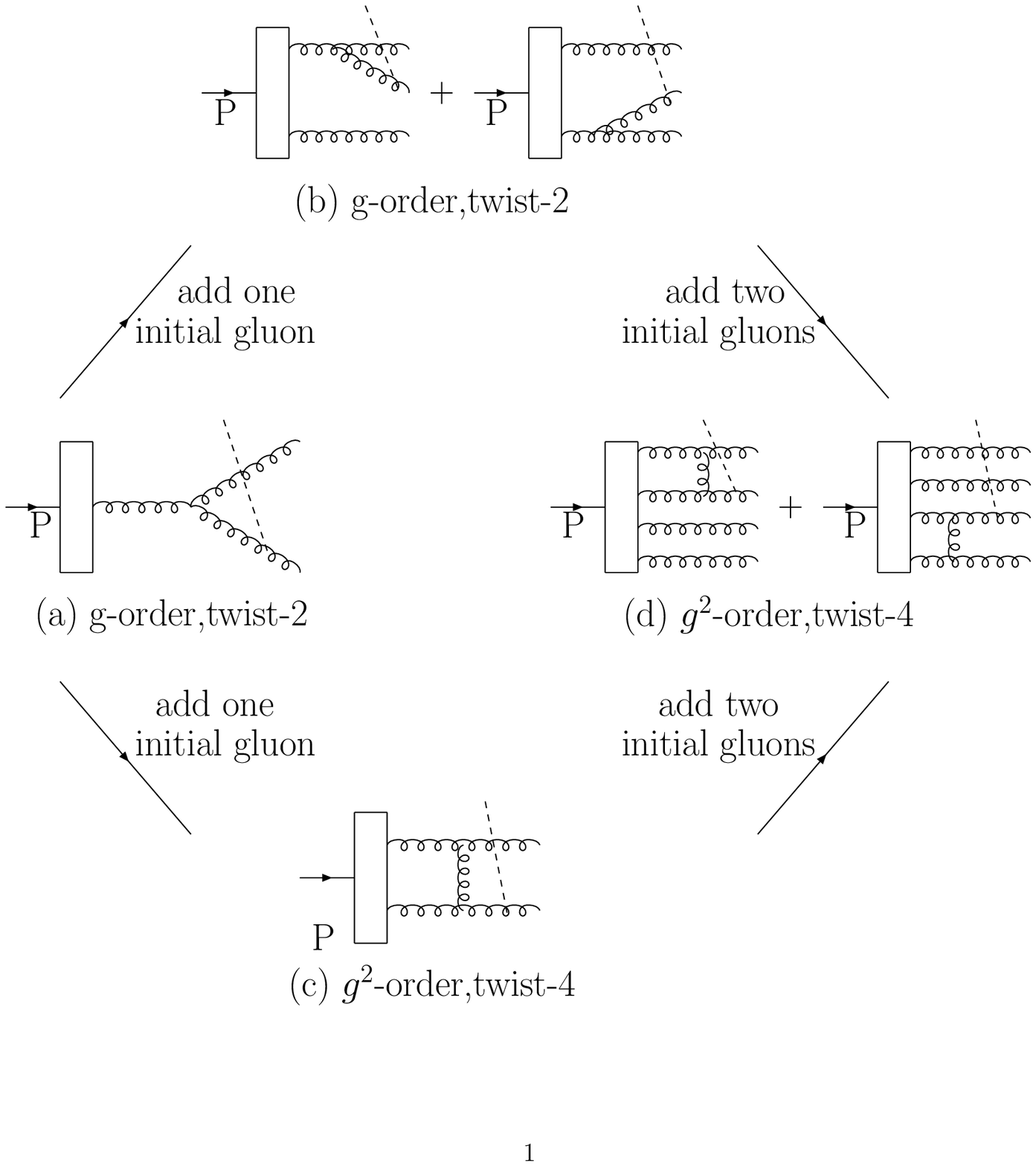,width=12.0cm,clip=}}}
 \vskip -4.0 truecm
Fig.1: The corrections of the initial gluons to a basic amplitude
of the DGLAP equation (a) and they lead to (b) the BFKL equation,
(c) the GLR-MQ-ZRS equation and (d) a new evolution equation,
respectively.  The dashed line is a virtual current which probing
gluon. Note that the four evolution equations form a closed
circuit, which implies a consistence among four evolution
equations.

\vskip 1.0 truecm

        We will present the derivations of the above mentioned four evolution equations
in a same partonic framework. For this sake, we use the Bjorken
frame, where the traditional parton distributions inside a fast
moving target are defined in the factorization scheme. Note that the
BFKL equation was originally derived by using the Regge langauge. In
this work we take an alternative technic to re-derive the BFKL
equation in Sec. 2, where the time ordered perturbation theory
(TOPT) [8] is used the same as the Altarelli-Parisi-derivation in
the DGLAP equation [2].

    The new derivation of the BFKL equation allows us conveniently to
add the corrections of the gluon fusion on it according to the
physical pictures in Fig. 1. We present the derivation of the new
evolution equation in Sec. 3. The nonlinear part of this equation
has IR divergences similar to the linear BFKL-kernel. Naturally, the
similar regularization scheme as in the BFKL equation is necessary.
Thus, we use the TOPT-cutting rule [4] to collect the contributions
from the virtual processes in the linear and nonlinear parts of the
new evolution equation. Four evolution equations at small $x$ in
Fig. 1 show their consistence. We discuss the relations among these
evolution equations in Sec. 4. We find that the new evolution
equation is a natural result following the DGLAP, BFKL, GLR-MQ-ZRS
and BK equations.

    Using the available saturation models as the input distribution,
we study the numerical solutions of Eq. (3.46) in Sec. 5. The solution
shows an unexpected result: the unintegrated gluon
distribution function $F(x,\underline{k}^2)$ in Eq. (3.46) begins
its smooth evolution under suppression of gluon recombination, but
when $x$ approaches a small critical value $x_c$,
$F(x,\underline{k}^2)$ will oscillate aperiodically in a narrow
$\underline{k}^2$ range (see Fig. 16). We find that this solution
presents the chaotic characteristics. In particular, this solution
of Eq. (3.46) has the positive Lyapunov exponents (Fig. 21), i.e.,
the solution is chaos.

    We indicate that chaos in Eq. (3.46) origins from a
serious of perturbations when $\underline{k}$ crosses over the
saturation scale. The rapid oscillation in chaos in a narrow
$\underline{k}^2$ domain arises a big shadowing (Fig. 15), which
blocks the QCD evolution vis three gluon vertex (Fig. 14). The chaos
effects in Eq. (3.46) are discussed in Sec. 6.

    Chaos, which has been observed in nature, is a
highlighted phenomenon in nonlinear physics. We proposed an example
where chaos appears in a QCD evolution equation and it may influence
the gluon distribution function, even change our expectations to the
future large hadron colliders.

    In this paper, sections 1-4 are the derivation of
the new evolution equation; sections 5-6 present the chaos solution
of this equation and its effects.

\newpage
\begin{center}
\section{The BFKL equation}
\end{center}

    We consider the following partonic picture of the DIS process.
At the lowest order, the elementary amplitude in Fig. 1a together
with its conjugate amplitude constructs the DGLAP equation for
gluon. However, this picture should be modified at small $x$ due to
the correlations among initial gluons. For example, a possible
correction to the DGLAP-amplitudes are given in Fig. 1b, or detailed
in Fig. 2. These processes imply that a scattered gluon is omitted
from two correlating gluons before its radiation. We call such a
correlated gluon cluster as the cold spot, which phenomenologically
describes the correlation among initial partons, where the dark
circle implies soft QCD-interactions. Neglecting the irrelevant part
with the evolution dynamics using the TOPT decomposition,
using the TOPT-decomposition Fig. 2 can been simplified as Fig. 3,
where the dashed lines are the time-ordering lines in the TOPT and
"x" marks the probing place. Note that the all lines across the time lines are
on mass-shell.

     The evolution kernel in QCD evolution equation is a part
of a complete scattering diagram. In general, the correlated initial
partons have the transverse momenta and they are off mass-shell,
therefore, the $\underline{k}$-factorization scheme is necessary. In
this work we use the semi-classical Weizs$\ddot{\rm a}$cker-Williams
($W-W$) approximation [9] to realize the
$\underline{k}$-factorization scheme. The $W-W$ approximation allows
us to extract the evolution kernels and to keep all initial and
final partons of the evolution kernels on their mass-shell.

\vskip -1.0 truecm \hbox{
\centerline{\epsfig{file=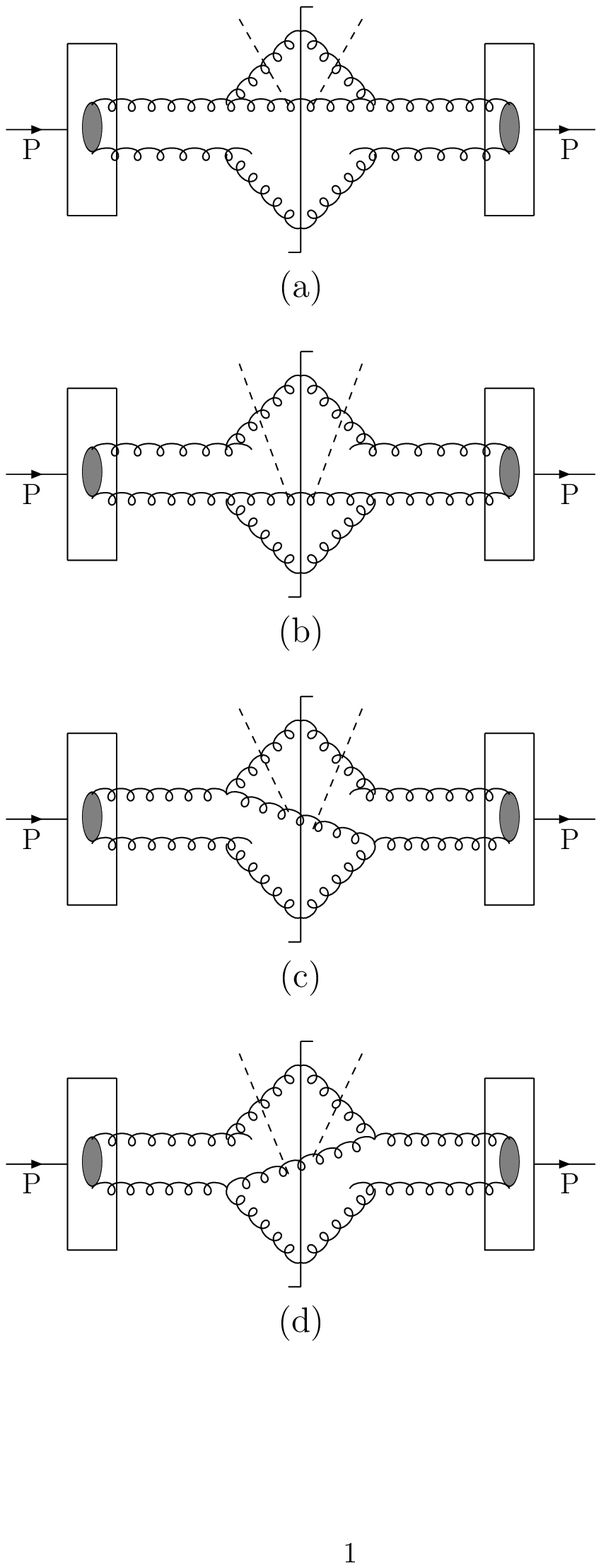,width=14.0cm,clip=}}}
 \vskip -4.0 truecm
\noindent Fig.2: The Feynman diagrams corresponding to the
elemental amplitudes of Fig. 1b. These diagrams lead to the real
part of the BFKL equation.

 \vskip 1.0 truecm

\vskip 0 truecm \hbox{
\centerline{\epsfig{file=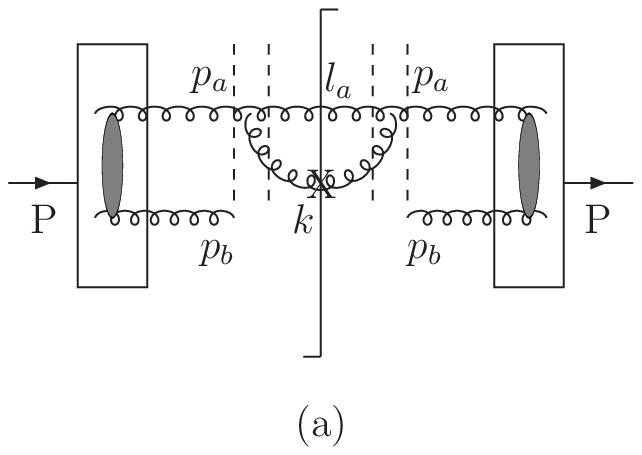,width=5.0cm,clip=}}}
\hbox{\centerline{\epsfig{file=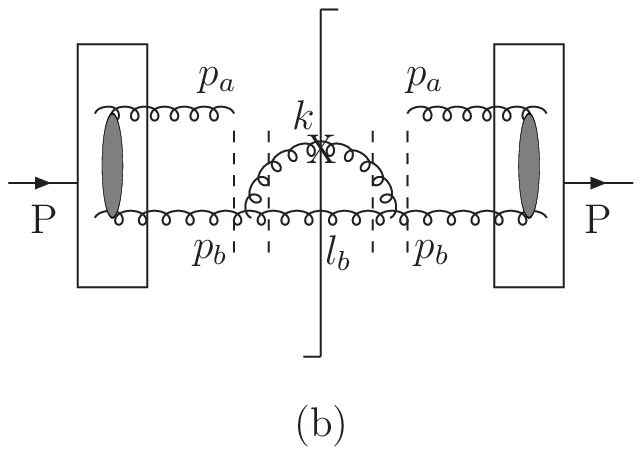,width=5.0cm,clip=}}}
\hbox{\centerline{\epsfig{file=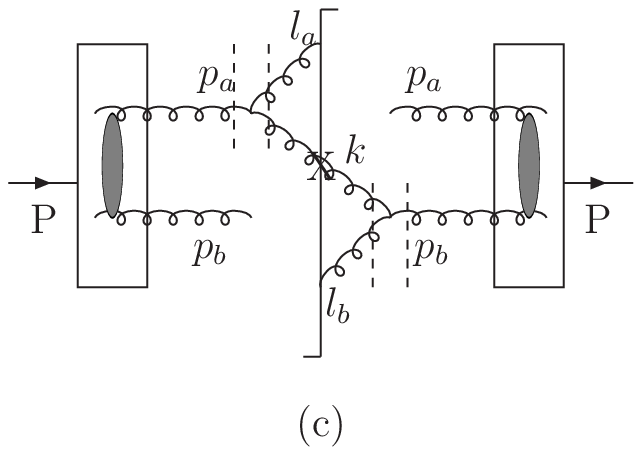,width=5.0cm,clip=}}}
\hbox{\centerline{\epsfig{file=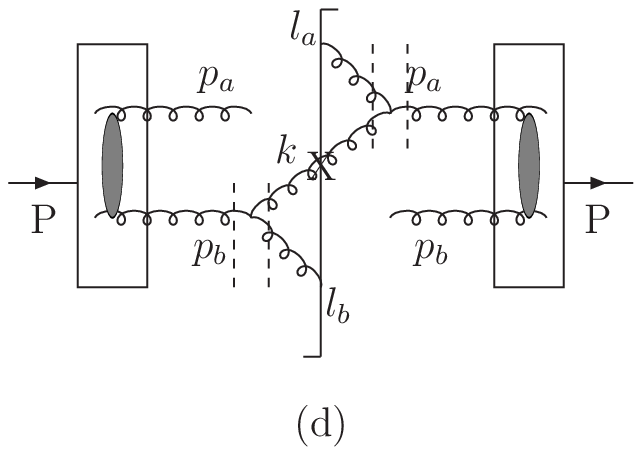,width=5.0cm,clip=}}}
 \vskip 0 truecm
Fig.3: The TOPT-diagrams corresponding to Fig. 2, the dashed lines
are the time ordered lines in the TOPT and "x" marks the probing
place. These diagrams lead to the real part of the BFKL equation.
For simplicity we neglect some parton lines, all those partons are
incorporated in the un-observed "X" state in the inclusive process
since they are irrelevant to the evolution kernel.

\vskip 1.0 truecm

According to the scale-invariant parton picture of the
renormalization group theory [10] the observed wave function
$\Psi(x_2,\underline{k})$ is evolved from the initial wave functions
$\Psi(x_1, \underline{p}_a)$ and $\Psi(x_1, \underline{p}_b)$ via
the QCD interactions, i.e.,

$$\Psi(x_2,\underline{k})=\Psi(x_1, \underline{p}_a)A_{BFKL1}+\Psi(x_1,
\underline{p}_b)A_{BFKL2}, \eqno(2.1)$$ where the two perturbative
amplitudes corresponding to Fig. 3 are

$$A_{BFKL1}=\sqrt{\frac{2E_{k}}{E_{p_a}+E_{p_b}}}\frac{1}{2E_{k}}
\frac{1}{E_{k}+E_{l_a}-E_{p_a}}M_1,\eqno(2.2)$$ and

$$A_{BFKL2}=\sqrt{\frac{2E_{k}}{E_{p_a}+E_{p_b}}}
\frac{1}{2E_{k}}\frac{1}{E_{k}+E_{l_b}-E_{p_b}}M_2.\eqno(2.3)$$

   The momenta of the partons are parameterized as

$$p_a=(x_1P+\frac{(\underline{k}+\underline{l}_a)^2}{2x_1P}
,\underline{k}+\underline{l}_a,x_1P), \eqno(2.4)$$

$$k=(x_2P+\frac{\underline{k}^2}{2x_2P},\underline{k},x_2P),\eqno(2.5)$$

$$l_a=((x_1-x_2)P+\frac{\underline{l}_a^2}{2(x_1-x_2)P},\underline{l}_a,
(x_1-x_2)P), \eqno(2.6)$$

$$p_b=(x_1P+\frac{(\underline{k}
+\underline{l}_b)^2} {2x_1P} ,\underline{k}+\underline{l}_b, x_1P),
\eqno(2.7)$$ and

$$l_b=((x_1-x_2)P+\frac{\underline{l}_b^2}
{2(x_1-x_2)P}, \underline{l}_b, (x_1-x_2)P). \eqno(2.8)$$

    The matrices of the local QCD interactions are

$$M_1=igf^{abc}[g_{\alpha\beta}(p_a+k)_{\gamma}+
g_{\beta\gamma}(-k+l_a)_{\alpha}+g_{\gamma\alpha}(-l_a-p_a)_{\beta}]
\epsilon_{\alpha}(p_a)\epsilon_{\beta}(k) \epsilon_{\gamma}(l_a),
\eqno(2.9)$$

$$M_2=igf^{abc}[g_{\alpha\beta}(p_b+k)_{\gamma}+
g_{\beta\gamma}(-k+l_b)_{\alpha}+g_{\gamma\alpha}(-l_b-p_b)_{\beta}]
\epsilon_{\alpha}(p_b)\epsilon_{\beta}(k) \epsilon_{\gamma}(l_b),
\eqno(2.10)$$ where the polarization vectors are

$$\epsilon(p_a)=(0,\underline{\epsilon},
-\frac{\underline{\epsilon}\cdot(\underline{k}+\underline{l}_a)}{x_1P}),
\eqno(2.11)$$

$$\epsilon(k)=(0,\underline{\epsilon},
-\frac{\underline{\epsilon}\cdot\underline{k}}{x_2P}),\eqno(2.12)$$
and

$$\epsilon(l_a)=(0,\underline{\epsilon},
-\frac{\underline{\epsilon}\cdot\underline{l}_a}{(x_1-x_2)P}),
\eqno(2.13)$$ where $\underline{\epsilon}$ is the transverse
polarization of the gluon in
$\epsilon_{\mu}=(\epsilon_0,\underline{\epsilon},\epsilon_3)=(0,\underline{\epsilon},0)$,
since the sum includes only physical transverse gluon states in the
TOPT form.

   Taking the $LL(1/x)$ approximation, i.e.,
assuming that $x_2\ll x_1$, one can get two similar amplitudes

$$A_{BFKL1}=igf^{abc}2\sqrt{\frac{x_1}{x_2}}
\frac{\underline{\epsilon}\cdot\underline{k}}{\underline{k}^2},
\eqno(2.14)$$ and

$$A_{BFKL2}=igf^{abc}2\sqrt{\frac{x_1}{x_2}}\frac{\underline{\epsilon}\cdot\underline{k}}{\underline{k}^2}.
\eqno(2.15)$$  However, these two amplitudes really occupy different
transverse configurations.  This is a reason why the dipole model of
the BFKL equation is derived by using the transverse
coordinator-space. However, we shall show that the momentum
representation still can be used to distinguish the differences
between Eqs. (2.14) and (2.15).

    The two parton correlation function is generally defined as

$$\vert
\Psi(x,\underline{p}_a,\underline{p}_b)\vert^2=f(x,\underline{p}_a,
\underline{p}_b)$$
$$=f\left(x,\frac{\underline{p}_a+\underline{p}_b}{2},\underline{p}_a-\underline{p}_b\right)
\equiv f(x,\underline{k}_c,\underline{k}_{ab}), \eqno(2.16) $$ where
$\underline{k}_c$ and $\underline{k}_{ab}$ are conjugate to the
impact parameter and transverse scale of a cold spot. Equation
(2.16) implies the probability of finding a gluon, which carries the
longitudinal momentum fraction $x$ of a nucleon and locates inside a
cold spot characterized by $\underline{k}_c$ and
$\underline{k}_{ab}$.

   In this work we derive the evolution equations in the impact parameter-independent case. This approximation
implies that the evolution dynamics of the partons are dominated
by the internal structure of the cold spot.  Thus, the evolution
kernel is irrelevant to $\underline{k}_c$ and we shall use

$$f(x,\underline{k}_{ab})=\int
\frac{d^2\underline{k}_c}{\underline{k}^2_c}f(x,\underline{k}_c,\underline{k}_{ab}),
\eqno(2.17)$$ which has the following TOPT-structure

$$f(x,\underline{k}_{ab})$$
$$\equiv \frac{E_{ab}}{2E_P}\vert M_{P\rightarrow k_{ab}X}\vert^2
\left[\frac{1}{E_P-E_{ab}-E_{X}}\right]^2\left[\frac{1}{2E_{ab}}\right]^2
\prod_X\frac{d^3k_X}{(2\pi)^32E_X}.
 \eqno(2.18)$$

    Notice that all transverse momenta in Eqs. (2.4)-(2.13) are
indicated relative to the mass-center of the nucleon target. However
according to Eq. (2.17), the evolution variable is the relative
momentum $\underline{k}_{ab}$, therefore, it is suitable to rewrite
all momenta relative to $\underline{p}_b$ in Eq. (2.2) and to
$\underline{p}_a$ in Eq. (2.3), respectively. Thus, we replace the
transverse momenta as follows:

$$\underline{p}_a\rightarrow \underline{p}_a-\underline{p}_b\equiv\underline{k}_{ab},$$
$$\underline{k}\rightarrow
\underline{k}-\underline{p}_b\equiv\underline{k}_{0b},$$ and
$$\underline{l}_a\rightarrow\underline{k}_{ab}-\underline{k}_{0b}=\underline{p}_a-\underline{k}\equiv
\underline{k}_{a0},\eqno(2.19)$$ in Eq. (2.2) since

$$\underline{k}_{ab}=\underline{k}_{a0}+\underline{k}_{0b},
\eqno(2.20)$$ and

$$\underline{p}_b\rightarrow \underline{p}_b-\underline{p}_a=\underline{k}_{ba},$$
$$\underline{k}\rightarrow
\underline{k}-\underline{p}_a\equiv\underline{k}_{0a},$$ and
$$\underline{l}_b\rightarrow \underline{k}_{ba}-\underline{k}_{0a}=\underline{p}_b-\underline{k}=\underline{k}_{b0},\eqno(2.21)$$
in Eq. (2.3). In consequence, we have

$$\Psi(x_1, \underline{p}_a)=\Psi(x_1, \underline{p}_b)=\Psi(x_1,
\underline{k}_{ab}), \eqno(2.22)$$ and

$$A_{BFKL}(\underline{k}_{a0},\underline{k}_{0b},x_1,x_2)= igf^{abc}2\sqrt{\frac{x_1}{x_2}}
\left[\frac{\underline{k}_{a0}}{\underline{k}_{a0}^2}+\frac{\underline{k}_{0b}}{\underline{k}_{0b}^2}
\right]\cdot\underline{\epsilon}, \eqno(2.23)$$ where we identify
two $\underline{\epsilon}$ in Eq. (2.23) since the measurements on
$(x_2,\underline{k}^2_{a0})$ and $(x_2,\underline{k}^2_{0b})$ are
really the same event.

     Equation (2.1) together with Eqs. (2.22) and (2.23) provide such a picture: a parent cold spot with the longitudinal momentum
fraction $x_1$ and transverse momentum $\underline{k}_{ab}$ radiates
a gluon, which has the longitudinal momentum fraction $x_2$ and the
transverse momentum $\underline{k}_{a0}$ (or $\underline{k}_{0b}$).
It is interesting that this is a picture like the dipole model but
in the full momentum space. In fact, using the Fourier
transformation, one can obtain the corresponding amplitude in the
dipole model [11]

$$A_{BFKL}(\underline{x}_{a0},\underline{x}_{0b},x_1,x_2)=
\int{\frac{d^2\underline{k}_{a0}d^2\underline{k}_{ob}}{(2\pi)^4}
A_{BFKL}(\underline{k}_{a0},\underline{k}_{0b}},x_1,x_2)e^{i\underline{k}_{a0}\cdot\underline{x}_{a0}
+i\underline{k}_{0b}\cdot\underline{x}_{0b}}$$
$$=igf^{abc}2\sqrt{\frac{x_1}{x_2}}[\frac{\underline{x}_{a0}}{\underline{x}_{a0}^2}+
\frac{\underline{x}_{0b}}{\underline{x}_{0b}^2}]\cdot\underline{\epsilon}.
\eqno(2.24)$$ where $\underline{x}$ is the conjugate coordinator
corresponding to the relative transverse momentum $\underline{k}$.

    We taking the square of the total amplitude,
one can get

$$d\sigma(q_{probe} P\rightarrow k'X)$$
$$=\frac{E_{ab}}{2E_P}\vert M_{P\rightarrow
k_{ab}X}\vert^2
\left[\frac{1}{E_P-E_{ab}-E_{X}}\right]^2\left[\frac{1}{2E_{ab}}\right]^2
\prod_X\frac{d^3k_X}{(2\pi)^32E_X}$$
$$\times \sum_{pol} A_{BFKL}A_{BFKL}^{\ast}\frac{
d^3k_{ab}}{(2\pi)^3E_{ab}}$$
$$\times \frac{1}{8E_kE_{probe}}\vert M_{q_{probe}k\rightarrow
k'}\vert^2(2\pi)^4\delta^4(q_{probe}+k-k')\frac{d^2k'}{(2\pi)^32E_{k'}}$$
$$=f(x_1,\underline{k}_{ab})\otimes\frac{x_1}{x_2}{\cal
K}_{BFKL}\left(\underline{k}_{ab},\underline{k}_{a0},\alpha_s\right)\otimes
d\sigma(q_{probe}^*k(x_2,\underline{k}_{a0})\rightarrow
k'(x_2,\underline{k}'))$$
$$\equiv \Delta
[\Psi(x_2,\underline{k}_{a0})\Psi^*(x_2,\underline{k}_{a0})+\Psi(x_2,\underline{k}_{a0})\Psi^*(x_2,\underline{k}_{0b})
+$$
$$\Psi(x_2,\underline{k}_{0b})\Psi^*(x_2,\underline{k}_{a0})+\Psi(x_2,\underline{k}_{0b})\Psi^*(x_2,\underline{k}_{0b})]\otimes
d\sigma(q_{probe}^*k(x_2,\underline{k}_{a0})\rightarrow
k'(x_2',\underline{k}'))$$
$$=\Delta f(x_2,\underline{k}_{a0})\otimes
d\sigma(q_{probe}^*k(x_2,\underline{k}_{a0})\rightarrow
k'(x_2,\underline{k}')), \eqno(2.25)$$ where the probe in the last
step only picks up the contributions from
$\Psi(x_2,\underline{k}_{a0})\Psi^*(x_2,\underline{k}_{a0})$, we
regard $\Delta f(x_2,\underline{k}_{a0})$ as the increment of the
distribution $f(x_1,\underline{k}_{ab})$ when it evolves from
$(x_1,\underline {k}_{ab})$ to $(x_2,\underline {k}_{a0})$.
Therefore we have

$$\Delta f(x_2,\underline{k}_{a0})$$
$$=\int\frac{d^2\underline{k}_{ab}}{\underline{k}^2_{ab}}\int^1_{x_{2}}\frac{dx_1}{x_1}
\frac{x_1}{x_2}{\cal
K}_{BFKL}\left(\underline{k}_{ab},\underline{k}_{a0},\alpha_s\right)
f(x_1,\underline{k}_{ab}), \eqno(2.26) $$ or

$$\Delta \tilde{F}(x_2,\underline{k}_{a0})\equiv \Delta x_2f(x_2,\underline{k}_{a0})$$
$$=\int\frac{d^2\underline{k}_{ab}}{\underline{k}^2_{ab}}\int^1_{x_{2}}\frac{dx_1}{x_1}
{\cal
K}_{BFKL}\left(\underline{k}_{ab},\underline{k}_{a0},\alpha_s\right)
\tilde{F}(x_1,\underline{k}_{ab}).  \eqno(2.27)$$  Using definition

$$\tilde{F}(x_2,\underline{k}_{a0})=\tilde{F}(x_1,\underline{k}_{ab})+\Delta
\tilde{F}(x_2,\underline{k}_{a0}), \eqno(2.28)$$ we write

$$-x\frac{\partial \tilde{F}(x,\underline{k}_{a0})}{\partial x}$$
$$=\int d^2\underline{k}_{ab}{\cal
K}_{BFKL}(\underline{k}_{ab},
\underline{k}_{a0},\alpha_s)\tilde{F}(x,\underline{k}_{ab}),
\eqno(2.29) $$

    According to Eq. (2.23), the evolution kernel reads as

$${\cal K}_{BFKL}(\underline{k}_{ab},\underline{k}_{a0},
\alpha_s)\frac{x_1}{x_2}\frac{dx_1}{x_1}=\sum_{pol}
A_{BFKL}A_{BFKL}^{\ast}\frac{dx_1}{2x_1}\frac{
1}{(2\pi)^3}$$
$$=\frac{\alpha_{s}N_c}{\pi^2}
\frac{\underline{k}_{ab}^2}{\underline{k}_{a0}^2\underline{k}_{0b}^2}\frac{dx_1}{x_2}.
\eqno(2.30)$$ Finally Eq. (2.29) becomes

$$-x\frac {\partial \tilde{F}(x,\underline{k}_{a0})}{\partial x}$$
$$=\frac{\alpha_{s}N_c}{\pi^2}\int d^2 \underline{k}_{ab}
\frac{\underline{k}_{ab}^2}{\underline{k}_{a0}^2\underline{k}_{0b}^2}
\tilde{F}(x,\underline{k}_{ab}). \eqno(2.31)$$  This is the real
part of the BFKL equation.

    The evolution kernel of the DGLAP
equation has infrared (IR) singularities, which relate to the
emission or absorption of quanta with zero momentum. A standard
regularized method is to combine the contributions of the
corresponding virtual processes. We call a cut diagram as the
virtual diagram, where one side of the cut line is a naive partonic
definition without any QCD corrections. A simple calculation of the
virtual diagrams was proposed via the TOPT cutting rule in [4]. Let
us summarize the TOPT cutting rule as follows. When we use a probe
to observe the parton distributions inside the target, we cannot
control the probing position. In principle, we should sum over all
cut diagrams belonging to the same time-ordered un-cut diagrams, and
these diagrams have similar singular structure but may come up with
opposite signs.  The TOPT-cutting rule presents the simple
connections among the related cut-diagrams including the real- and
virtual-diagrams. The BFKL-kernel also has singularities on the
transverse momentum space. Thus, we can pick up the contributions
from the virtual diagrams using the TOPT-cutting rule without the
complicated calculations.

\vskip 0.5 truecm \hbox{
\centerline{\epsfig{file=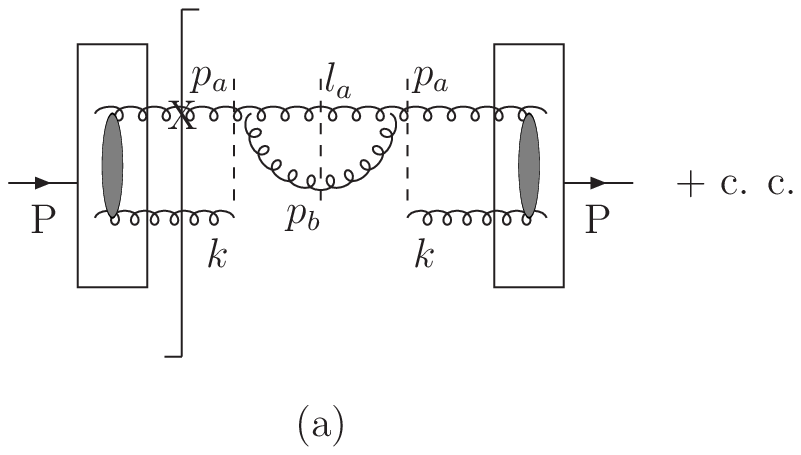,width=6.0cm,clip=}}} \hbox{
\centerline{\epsfig{file=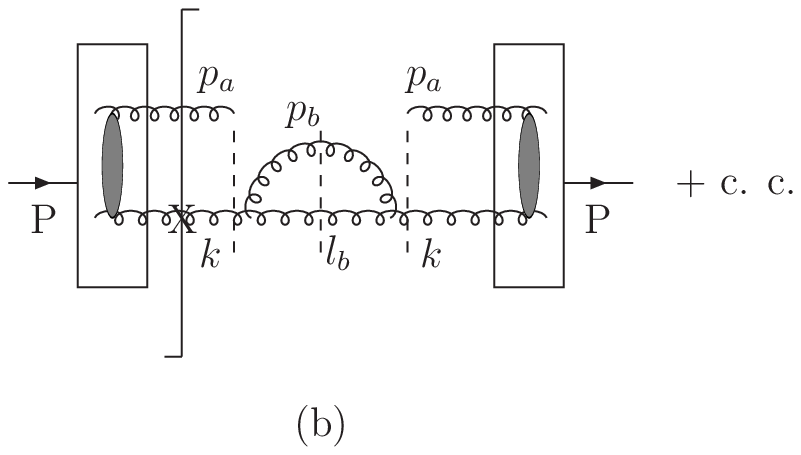,width=6.0cm,clip=}}} \hbox{
\centerline{\epsfig{file=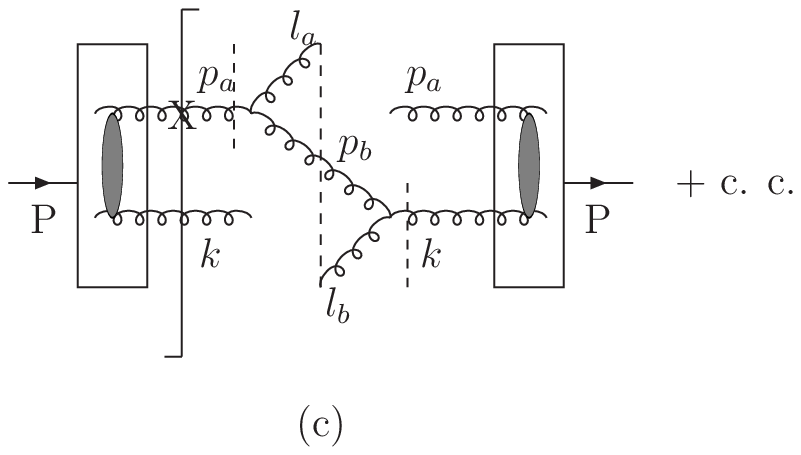,width=6.0cm,clip=}}} \hbox{
\centerline{\epsfig{file=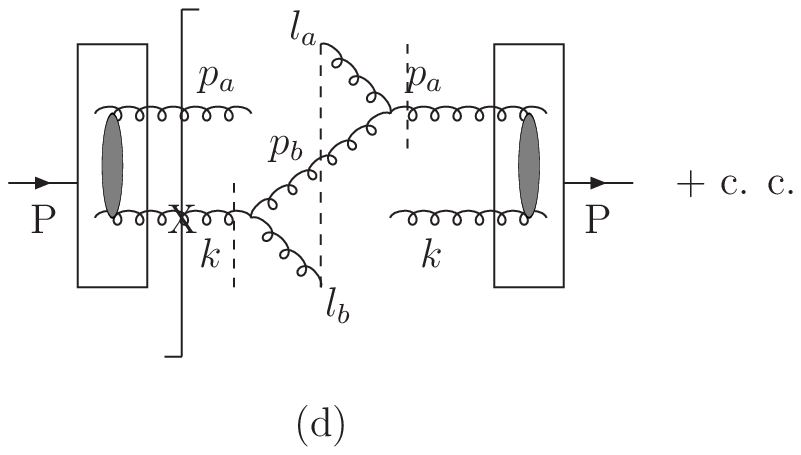,width=6.0cm,clip=}}}
 \vskip 0 truecm

 Fig.4: We call these figures and their conjugate figures
as the virtual diagrams corresponding to Fig. 3.

 \vskip 1.0 truecm

    Using the TOPT-cutting rule, one can prove that the
diagrams in Fig. 4 contribute a similar evolution kernel as the real
kernel but differ by a factor $-1/2\times(1/2+1/2)$. The negative
sign arises from the changes of time order in the energy
denominators. The factor $(1/2+1/2)$ is due to the fact that the
probe ``sees'' only the square root of the parton distribution,
which accepts the contributions of the partonic processes in a
virtual diagram, and the other factor $1/2$ is originated from the
symmetry of the pure gluon process. Therefore, the evolution
equation corresponding to Fig. 4 is

$$-x\frac {\partial \tilde{F}(x,\underline{k}_{ab})}{\partial x}$$
$$=-\frac{1}{2}\frac{\alpha_{s}N_c}{\pi^2}\int d^2 \underline{k}_{a0}
\frac{\underline{k}_{ab}^2}{\underline{k}_{a0}^2(\underline{k}_{ab}-\underline{k}_{a0})^2}
\tilde{F}(x,\underline{k}_{ab}). \eqno(2.32)$$  Since we calculate
the contributions to $\Delta\tilde{F}(x,\underline{k}_{a0})$, we
should make the replacement $b\leftrightarrow 0$ in Eq. (2.32).
Combining the real and virtual parts of the evolution equation, we
have

$$-x\frac {\partial \tilde{F}(x,\underline{k}_{a0})}{\partial x}$$
$$=\frac{\alpha_{s}N_c}{2\pi^2}\int d^2
\underline{k}_{ab}\left[2\frac{\underline{k}_{ab}^2}{\underline{k}_{a0}^2\underline{k}_{0b}^2}\tilde{F}(x,\underline{k}_{ab})
-\frac{\underline{k}^2_{a0}}{\underline{k}^2_{ab}\underline{k}^2_{0b}}\tilde{F}(x,\underline{k}_{a0})\right].
\eqno(2.33)$$

   According to Eq. (2.18), the distribution $ f(x,\underline{k})$
in the TOPT-form contains a singular factor $1/\underline{k}^4$,
which arises from the off energy-shell effect in the square of the
energy denominator. In order to ensure the safety using of the $W-W$
approximation, we move this factor to the evolution kernel and use
the following new definition of the unintegrated gluon distribution

$$F(x,\underline{k})=\frac{\underline{k}^4}{\hat{\underline{k}}^4}
\tilde{F}(x,\underline{k}),  \eqno(2.34)$$ where
$\hat{\underline{k}}$ is a unity vector on the transverse momentum
space.  Thus, Eq. (2.33) becomes

$$-x\frac {\partial F(x,\underline{k}_{a0})}{\partial x}$$
$$=\frac{\alpha_{s}N_c}{2\pi^2}\int d^2
\underline{k}_{ab}
\frac{\underline{k}_{a0}^2}{\underline{k}_{ab}^2\underline{k}_{0b}^2}
\left[2F(x,\underline{k}_{ab})-F(x,\underline{k}_{a0})\right],
\eqno(2.35)$$ which is consistent with a standard form of the BFKL
equation.

   The correlations among the initial
gluons can be neglected in the dilute parton system. In this case
the contributions of the interference diagrams Figs.3c and 3d
disappear. Thus, the kernel Eq. (2.30) reduces to the splitting
functions in the DGLAP equation at the small $x$ limit, i.e.,

$${\cal K}_{BFKL}(\underline{k}_{ab},\underline{k}_{a0},
\alpha_s)\frac{x_1}{x_2}\frac{dx_1}{x_1}d^2\underline{k}_{ab}\rightarrow
\frac{\alpha_sN_c}{\pi}
\frac{dx_1}{x_2}\frac{d\underline{k}^2}{\underline{k}^2}$$
$$\equiv {\cal
K}_{DGLAP}\frac{d\underline{k}^2}{\underline{k}^2}\frac{dx_1}{x_1}.\eqno(2.36)$$
Since in this case two initial gluons have the same transverse
momentum, we can always take it to zero and use the collinear
factorization to separate the gluon distribution. The corresponding
DGLAP equation reads

$$Q^2\frac{\partial g(x_B,Q^2)}{\partial Q^2} =\int_{x}^1\frac{dx_1}{x_1}
{\cal K}_{DGLAP}\left(\frac{x_B}{x_1},\alpha_s\right)g(x_1,Q^2)$$
$$=\frac{\alpha_sN_c}{\pi}\int_{x_B}^1\frac{dx_1}{x_1}
\frac{x_1}{x_B} g(x_1,Q^2), \eqno(2.37)$$ where the scaling
restriction $\delta(x_2-x_B)$ is included and

$$G(x,Q^2)\equiv
xg(x,Q^2)=\int^{Q^2}_{\underline{k}^2_{min}}\frac{d\underline{k}^2}{\underline{k}^2}
xf(x,\underline{k}^2)\equiv\int^{Q^2}_{\underline{k}^2_{min}}\frac{d\underline{k}^2}{\underline{k}^2}
F(x,\underline{k}^2)\equiv
\int^{Q^2}_{\underline{k}^2_{min}}d\underline{k}^2
{\mathcal{F}}(x,\underline{k}^2). \eqno(2.38)$$

\newpage
\begin{center}
\section{The new evolution equation}
\end{center}

        We consider the evolution kernel based on Fig. 1d, which constructs a new evolution equation.
Notice that the two pairs of initial gluons, which are hidden in the
correlation function, for example in Fig. 5a, should be indicated as
Fig. 5b.

\vskip 0.5 truecm \hbox{
\centerline{\epsfig{file=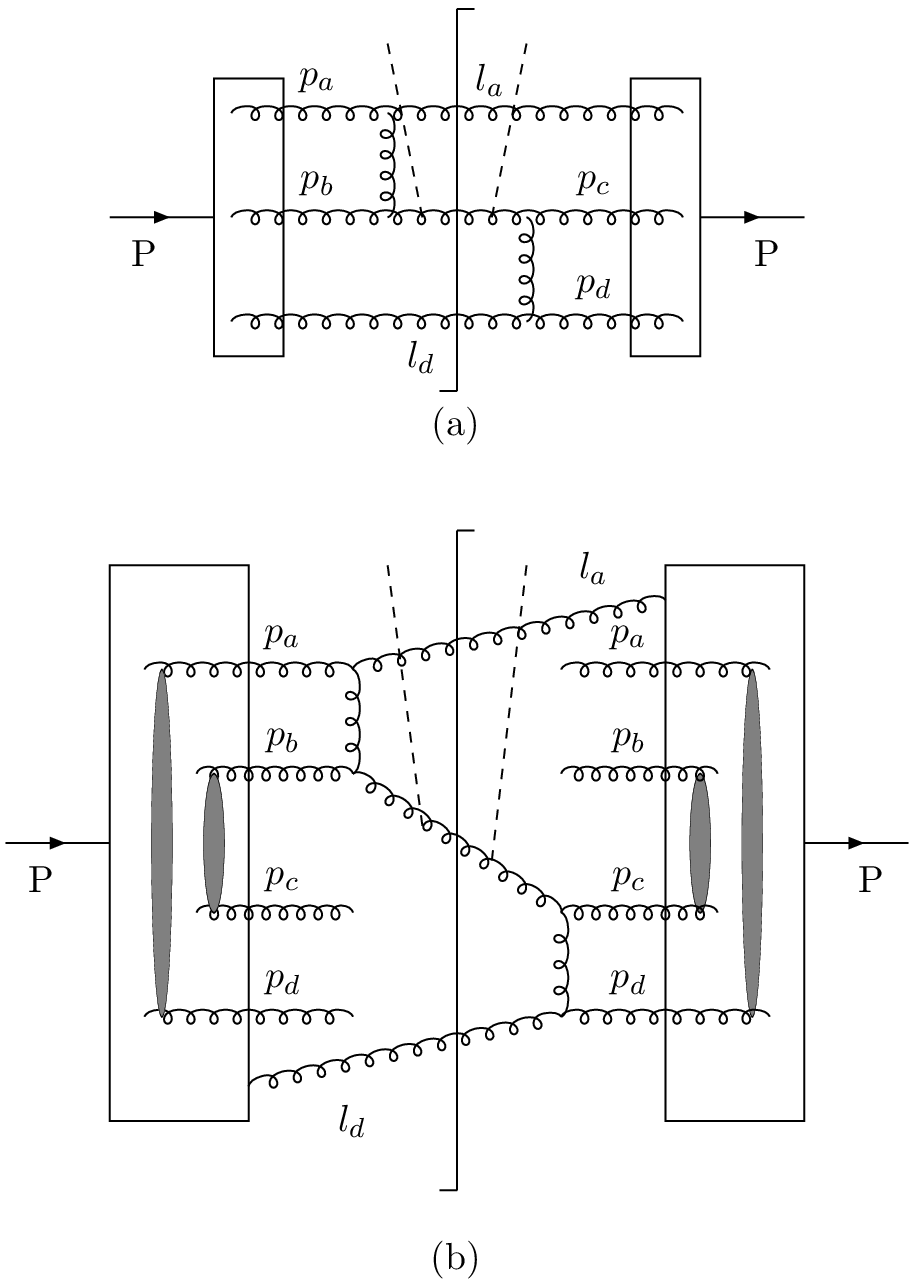,width=6.0cm,clip=}}}
 \vskip 0 truecm

 Fig.5: A cutting diagram originating from Fig. 1d.  For
simplicity we neglect some parton lines linking with $p_a$, $p_b$,
$p_c$ and $p_d$, all those partons are incorporated in the
un-observed "X" state in the inclusive process.

 \vskip 1.0 truecm

    A set of cut diagrams based on Fig. 1d are listed in Fig.
6, where the probe vertex has been separated out using the $W-W$
approximation and its position is indicated by "x".

    Similar to the derivation of Eq. (2.30), we write the
evolution kernel of the new evolution equation as

$${\cal K}_{New}=\frac{1}{16\pi^2}\frac{x_2}{x_1}\sum_{pol}
A_{New}A_{New}^{\ast}. \eqno(3.1)$$

    The amplitude

$$A_{New}=A_{New1}+A_{New2}, \eqno(3.2)$$
where

$$A_{New1}=\sqrt{\frac{2E_k}{E_{p_a}+E_{p_b}}}\frac{1}{2E_k}
\frac{1}{E_k+E_{l_a}-E_{p_a}-E_{p_b}}M_{New1},\eqno(3.3)$$ and

$$A_{New2}=\sqrt{\frac{2E_k}{E_{p_c}+E_{p_d}}}
\frac{1}{2E_k} \frac{1}{E_k+E_{l_d}-E_{p_c} -E_{p_d}}M_{New2}.
\eqno(3.4)$$

\vskip 0.5 truecm \hbox{
\centerline{\epsfig{file=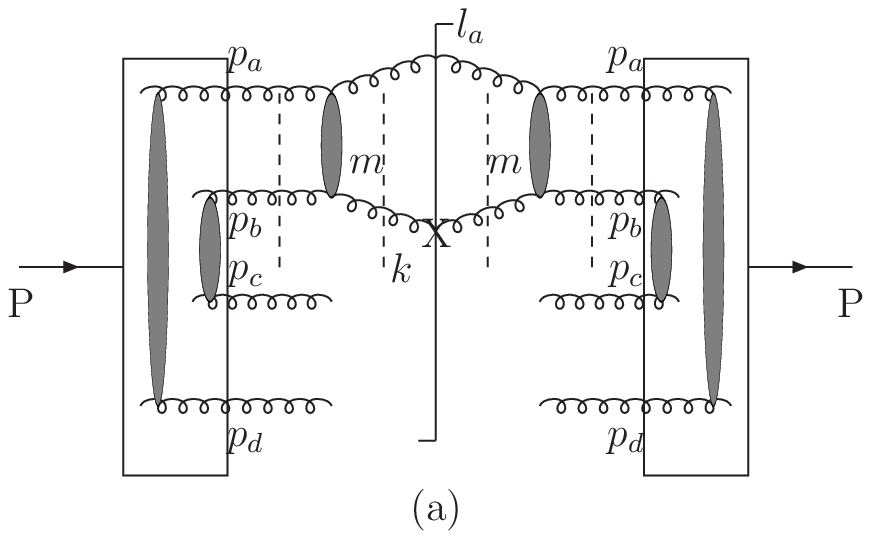,width=5.5cm,clip=}}} \hbox{
\centerline{\epsfig{file=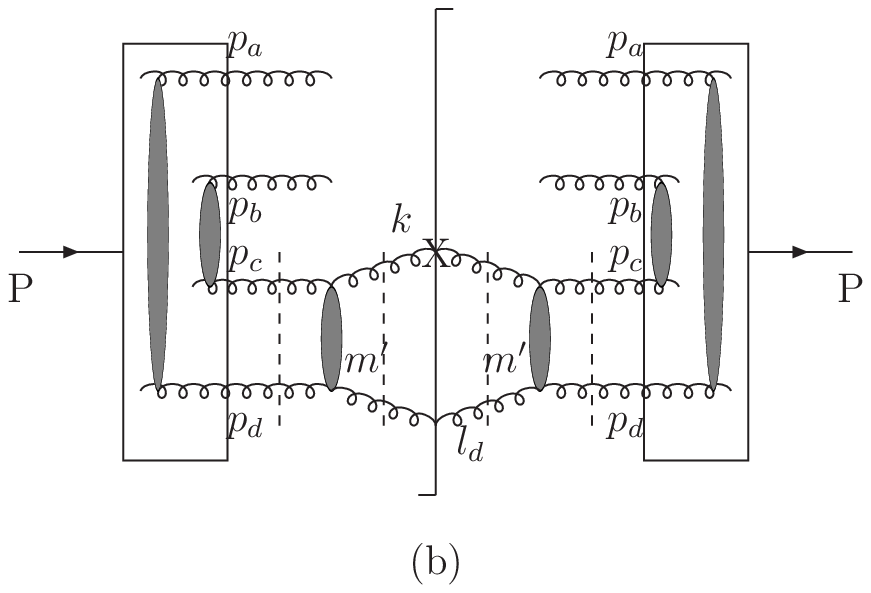,width=5.5cm,clip=}}} \hbox{
\centerline{\epsfig{file=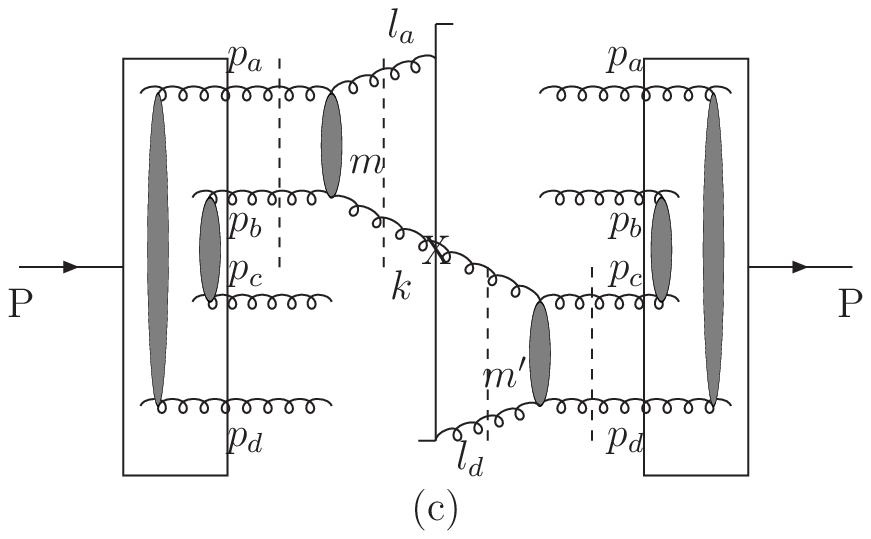,width=5.5cm,clip=}}} \hbox{
\centerline{\epsfig{file=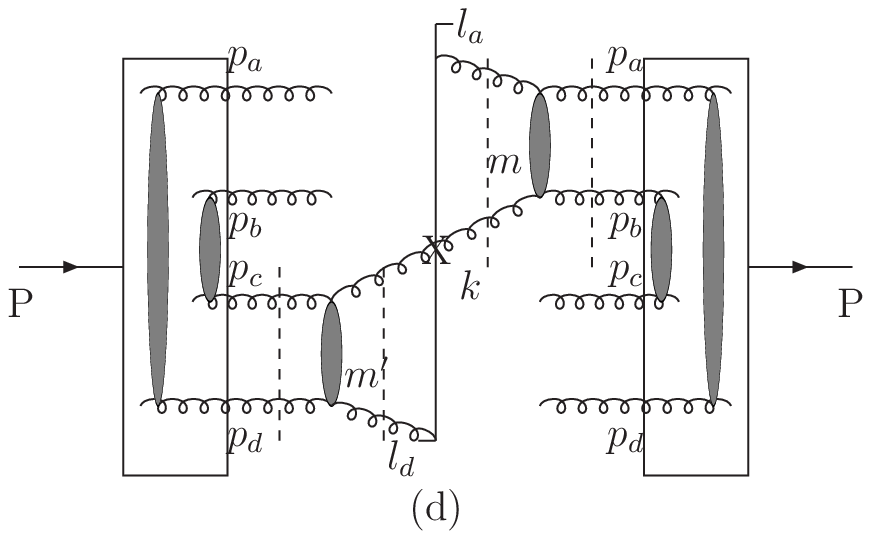,width=5.5cm,clip=}}}
 \vskip 0 truecm

 Fig.6: The TOPT-diagrams constructed by the elemental
amplitudes in Fig. 1d. For simplicity we neglect some lines
linking with $l_a$, $l_b$... in 7c and 7d, since they are
irrelevant to the evolution kernel.

 \vskip 1.0 truecm

\vskip 0.5 truecm \hbox{
\centerline{\epsfig{file=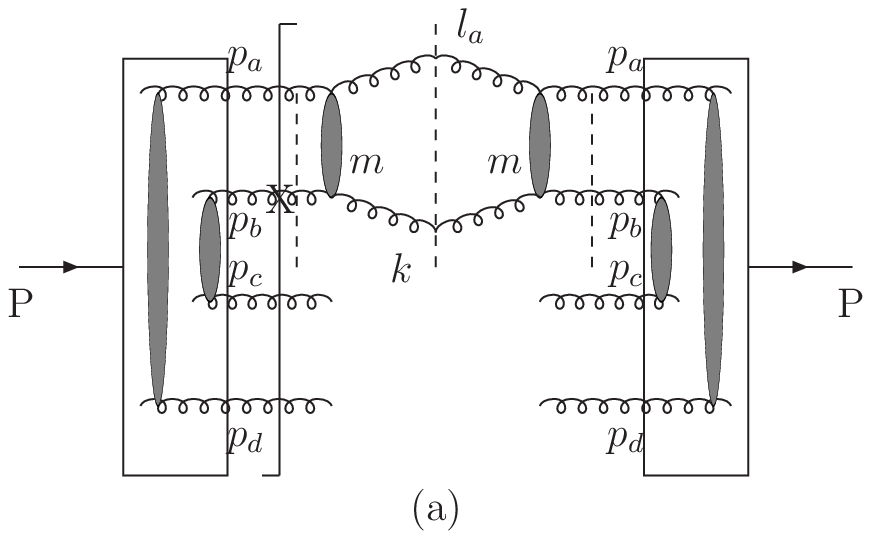,width=5.5cm,clip=}}} \hbox{
\centerline{\epsfig{file=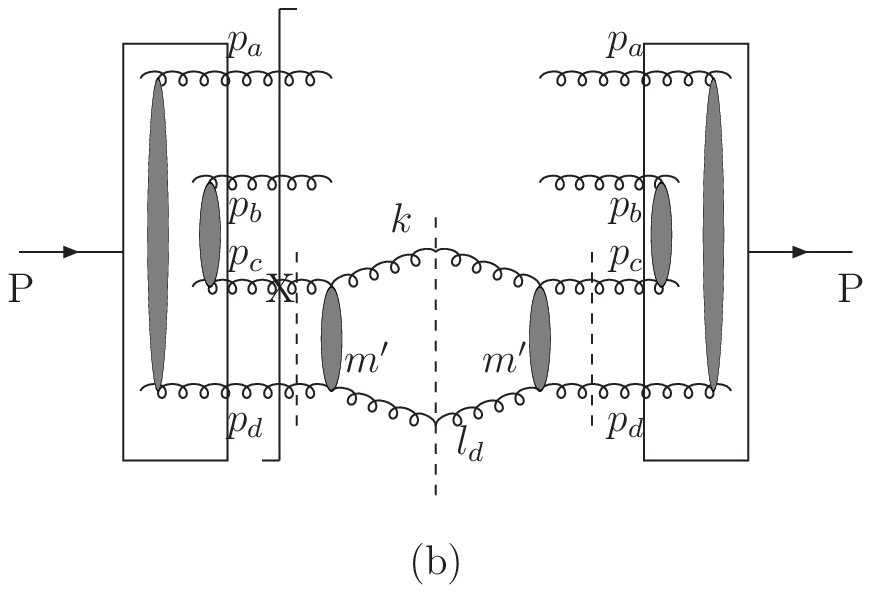,width=5.5cm,clip=}}} \hbox{
\centerline{\epsfig{file=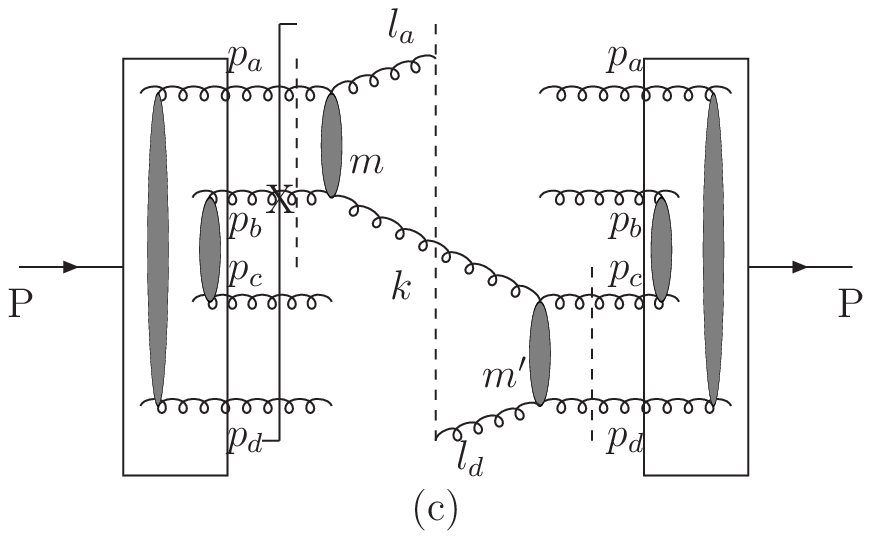,width=5.5cm,clip=}}} \hbox{
\centerline{\epsfig{file=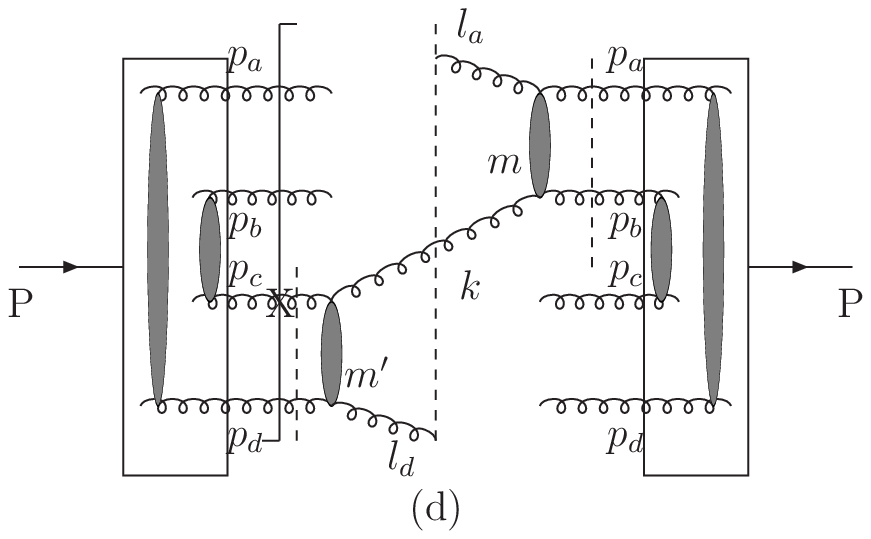,width=5.5cm,clip=}}}
 \vskip 0 truecm

 Fig.7: The virtual diagrams corresponding to Fig. 6.

 \vskip 1.0 truecm

 The momenta of the partons, for example, are parameterized as

$$p_a=(x_1P+\frac{(\underline{l}_a-\underline{m})^2}{2x_1P}
,\underline{l}_a-\underline{m},x_1P), \eqno(3.5)$$

$$p_b=(x_1P+\frac{(\underline{k}+\underline{m})^2}{2x_1P}
,\underline{k}+\underline{m},x_1P), \eqno(3.6)$$

$$k=(x_2P+\frac{\underline{k}^2}{2x_2P},\underline{k},x_2P),\eqno(3.7)$$

$$l_a=((2x_1-x_2)P+\frac{\underline{l}_a^2}{2(2x_1-x_2)P},\underline{l}_a,
(2x_1-x_2)P). \eqno(3.8)$$

$$p_c=(x_1P+\frac{(\underline{k}+\underline{m}')^2}{2x_1P}
,\underline{k}+\underline{m}',x_1P), \eqno(3.9)$$

$$p_d=(x_1P+\frac{(\underline{l}_a-\underline{m}')^2}{2x_1P}
,\underline{l}_a-\underline{m}',x_1P), \eqno(3.10)$$

$$l_d=((2x_1-x_2)P+\frac{\underline{l}_d^2}{2(2x_1-x_2)P},\underline{l}_d,
(2x_1-x_2)P). \eqno(3.11)$$ For example, in the t-channel

$$m=p_b-k=((x_1-x_2)P+\frac{(\underline{k}+\underline{m})^2}{2x_1P}
-\frac{\underline{k}^2}{2x_2P},\underline{m},(x_1-x_2)P),
\eqno(3.12)$$ and

$$m'=p_c-k=((x_1-x_2)P+\frac{(\underline{k}+\underline{m}')^2}{2x_1P}
-\frac{\underline{k}^2}{2x_2P},\underline{m}',(x_1-x_2)P).
\eqno(3.13)$$

    The matrices in Eqs. (3.3) and (3.4) are

$$M_{New1}=igf^{ABC}C^{\alpha\beta\gamma}\frac{-id^{\gamma\eta}_{\perp}}
{m^2}igf^{dce}C^{\rho\sigma\eta}\epsilon_{\alpha}(p_a)
\epsilon_{\rho}(p_b)\epsilon_{\beta}^{\ast}(l_a)
\epsilon^{\ast}_{\sigma}(k),\eqno(3.14) $$ and

$$M_{New2}=igf^{ABC}C^{\alpha\beta\gamma}\frac{-id^{\gamma\eta}_{\perp}}
{m^2}igf^{dce}C^{\rho\sigma\eta}\epsilon_{\alpha}(p_d)
\epsilon_{\rho}(p_c)\epsilon_{\beta}^{\ast}(l_d)
\epsilon^{\ast}_{\sigma}(k), \eqno(3.15)$$ where
$d^{\gamma\eta}_{\perp}=\overline{n}^{\gamma}n^{\eta}+\overline{n}^{\eta}n^{\gamma}-g^{\gamma\eta}$,
$C^{\alpha\beta\gamma}C^{\rho\sigma\eta}$ are the triple gluon
vertices and the polarization vectors are

$$\epsilon(p_a)=(0,\underline{\epsilon},
-\frac{\underline{\epsilon}\cdot(\underline{l}_a-\underline{m})}{x_1P}),\eqno(3.16)$$

$$\epsilon(p_b)=(0,\underline{\epsilon},
-\frac{\underline{\epsilon}\cdot(\underline{k}+\underline{m})}{x_1P}),\eqno(3.17)$$

$$\epsilon(k)=(0,\underline{\epsilon},
-\frac{\underline{\epsilon}\cdot\underline{k}}{x_2P}),
\eqno(3.18)$$ and

$$\epsilon(l_a)=(0,\underline{\epsilon},
-\frac{\underline{\epsilon}\cdot\underline{l}_a}{(2x_1-x_2)P}).
\eqno(3.19)$$ Thus, at small $x$ we have

$$A_{New}(\underline{k},x_1,x_2)$$
$$=g^2f^{ABC}f^{DCE}\sqrt{\frac{x_1}{2x_2}}
\left[6\frac{\underline{\epsilon}\cdot\underline{k}}{\underline{k}^2}
\frac{\underline{\epsilon}\cdot\underline{k}}{\underline{k}^2}
+6\frac{\underline{\epsilon}\cdot\underline{k}}{\underline{k}^2}
\frac{\underline{\epsilon}\cdot\underline{k}}
{\underline{k}^2}\right], \eqno(3.20)$$ where one of the two factors in
each term is from the approximation

$$\epsilon(k)m/m^2\simeq
\underline{\epsilon}\cdot\underline{k}/\underline{k}^2,$$ and

$$\epsilon(k)m'/m'^2\simeq
\underline{\epsilon}\cdot\underline{k}/\underline{k}^2.\eqno(3.21)$$

     We use
the relative transverse momenta to replace the relating momenta in
Eqs. (3.5)-(3.13) and recalculate Eqs. (3.3) and (3.4). The result
is

$$A_{New}(\underline{k},x_1,x_2)$$
$$= g^2f^{ABC}f^{DCE}\sqrt{\frac{x_1}{2x_2}}
\left[6\frac{\underline{\epsilon}\cdot\underline{k}_{(p_b,p_c)or(p_a,p_d)}(m)}{\underline{k}^2_{(p_b,p_c)or(p_a,p_d)}(m)}~
\frac{\underline{\epsilon}\cdot\underline{k}_{(p_b,p_c)}}{\underline{k}^2_{(p_b,p_c)}}\right.$$
$$\left.+6\frac{\underline{\epsilon}\cdot\underline{k}_{(p_b,p_c)or(p_a,p_d)}(m')}{\underline{k}^2_{(p_b,p_c)or(p_a,p_d)}
(m')}~ \frac{\underline{\epsilon}\cdot\underline{k}_{(p_b,p_c)}}{
\underline{k}^2_{(p_b,p_c)}}\right], \eqno(3.22)$$ where the
foot-indexes of the relative transverse momenta indicate the
corresponding cold spots and $\underline k(m)$, $\underline k(m')$
imply that the momenta origin from $m$, $m'$, respectively.  Using
the definitions

$$\underline{k}_{bc}=\underline{p}_b-\underline{p}_c,~~\underline{k}_{b0}=\underline{p}_b-\underline{k},~~
\underline{k}_{0c}=\underline{k}-\underline{p}_c, \eqno(3.23)$$ we
have

$$\underline{k}_{bc}=\underline{k}_{b0}+\underline{k}_{0c}.
\eqno(3.24)$$ We read two momenta $\underline{k}_{(p_b,p_c)}$ in
Eq. (3.22) as $\underline{k}_{0c}$ and $\underline{k}_{b0}$,
respectively. On the other hand, due to momentum conservation, we
have

$$\underline{k}_{(p_b,p_c)}(m)\equiv
\underline{p}_b-\underline{k}=\underline{p}_b-\underline{p}_c-\underline{k}+\underline{p}_c
=\underline{k}_{bc}-\underline{k}_{oc}=\underline{k}_{b0},$$  and

$$\underline{k}_{(p_b,p_c)}(m')\equiv\underline{k}-\underline{p}_c=
\underline{k}-\underline{p}_b-\underline{p}_c+\underline{p}_b
=\underline{k}_{0b}-\underline{k}_{cb}=\underline{k}_{0c}.
\eqno(3.25)$$ Thus, we obtain

$$A_{New}$$
$$=12g^2f^{ABC}f^{DCE}\sqrt{\frac{x_1}{2x_2}}
\frac{\underline{\epsilon}\cdot\underline{k}_{b0}
\underline{\epsilon}\cdot\underline{k}_{0c}}{\underline{k}_{b0}^2\underline{k}_{0c}^2}.
\eqno(3.26)$$ Note that the two factors $\underline{k}_{b0}^2$ and
$\underline{k}_{0c}^2$ in the denominator of Eq. (3.26) are
correlated through Eq. (3.24) and they have double poles as in the
BFKL-kernel (2.34).

    The result Eq.(3.26) seems irrelevant to
$\underline{p}_a$ and $\underline{p}_d$. However, there are two
possible contributions of the cold spot $(p_a,p_d)$ to the
evolution kernel:

    (1) The momenta $\underline{p}_a$ and $\underline{p}_d$ don't flow
into the amplitude Eq. (3.26). Therefore, the cold spot $(p_a, p_d)$
in Fig. 6b is independent of the evolution dynamics and its
distribution should be integrated as a unobservable quantity. Thus,
the resulting kernel reduces to the linear BFKL kernel.

    (2) The momenta $\underline{p}_a$ and $\underline{p}_d$ flow
into the amplitude Eq. (3.26) through $\underline{m}$ and
$\underline{m}'$. The momenta $\underline{k}_{(p_b,p_c)}(m)$ and
$\underline{k}_{(p_b,p_c)}(m')$ in Eq. (3.22) are alternatively
replaced by $\underline{k}_{(p_a,p_d)}(m)=
\underline{p}_a-\underline{k}\equiv\underline{k}_{a0}$ and
$\underline{k}_{(p_a,p_d)}(m')=\underline{k}-\underline{p}_d\equiv\underline{k}_{0d}$,
respectively. The corresponding amplitudes become

$$A'_{New}$$
$$=6g^2f^{ABC}f^{DCE}\sqrt{\frac{x_1}{2x_2}}
\frac{\underline{\epsilon}\cdot\underline{k}_{a0}
\underline{\epsilon}\cdot\underline{k}_{0c}}{\underline{k}_{a0}^2\underline{k}_{0c}^2},
\eqno(3.27)$$ and

$$A''_{New}$$
$$=6g^2f^{ABC}f^{DCE}\sqrt{\frac{x_1}{2x_2}}
\frac{\underline{\epsilon}\cdot\underline{k}_{0d}
\underline{\epsilon}\cdot\underline{k}_{b0}}{\underline{k}_{0d}^2\underline{k}_{b0}^2},
\eqno(3.28)$$ where one can introduce

$$\underline{k}_{ab}\equiv
\underline{p}_a-\underline{p}_d=\underline{p}_a-\underline{k}-\underline{p}_d+\underline{k}
=\underline{k}_{a0}+\underline{k}_{0d}.\eqno(3.29)$$ In general,
the momenta $\underline{k}_{a0}$ and $\underline{k}_{0d}$ in Eqs.
(3.27) and (3.28) are undetermined since $l_a$ and $l_d$ in Fig.
6b are unobserved, they should be integrated out as two
independent variables. Thus, the resulting evolution kernel reduce
to the DGLAP-like kernel.

      Obviously, the above mentioned two situations should be excluded in our resummation
in order to get the leading corrections, unless we have the following
restriction conditions

$$\underline{k}_{a0}=\underline{k}_{b0},$$

$$\underline{k}_{0d}=\underline{k}_{0c}, \eqno(3.30)$$ and they
imply that

$$\underline{k}_{ad}=\underline{k}_{bc}, \eqno(3.31)$$ due to Eqs. (3.24) and (3.29).
To understand Eq. (3.31), we image that before the probe interacts
with the target, two overlapping cold spots have recombined into a
common cold spot $(\underline{p}_b,\underline{p}_c)$,
This is an inverse processes of the dipole splitting in
the BK equation [5]. Therefore the
probe always measures the recombination processes of four initial
gluons originated from a same cold spot and sharing a same relative
momentum.

    Summing all the channels, we get the evolution kernel
corresponding to Fig. 6 and the result reads

$${\cal K}_{New}\frac{x_1}{x_2}\frac{dx_1}{x_1}d^2\underline{k}_{bc}$$
$$=\sum_{pol}
A_{New}A_{New}^{\ast}\left[\frac{1}{16\pi^3}\frac{dx_1}{x_1}d^2\underline{k}_{bc}\right]$$
$$=\frac{9\alpha^2_s}{2\pi}\frac{N_c^2}{N_c^2-1}\frac{1
}{\underline{k}_{bc}^2}
\frac{\underline{k}_{bc}^2}{\underline{k}_{b0}^2\underline{k}_{c0}^2}\frac{dx_1}{x_2}d^2\underline{k}_{bc}.
\eqno(3.32)$$

    In the case of decreasing gluon density, the
contributions of the interference terms (Figs. 7c and 7d) disappear
and Fig. 1d return to Fig. 1c. Thus, Eq. (3.32) reduces to the real
part of the GLR-MQ-ZRS kernel [12]

$${\cal K}_{New}\frac{x_1}{x_2}\frac{dx_1}{x_1}d^2\underline{k}_{bc}
\rightarrow \frac{9\alpha^2_s}{2\pi}
\frac{N_c^2}{N_c^2-1}\frac{dx_1}{x_2}\frac{d^2\underline{k}}{\underline{k}^4}$$
$$\equiv {\cal K}_{GLR-MQ-ZRS}\frac{dx_1}{x_1}\frac{d\underline{k}^2}{\underline{k}^4}. \eqno(3.33)$$

      Thus, we have

$$G(x_2,Q^2_2)=G(x_1,Q^2_1)+\Delta G(x_2,Q^2_2)$$
$$=G(x_1,Q^2_1)+\int^{Q^2_2}_{Q^2_{1min}}\frac{dQ^2_1}{Q^4_1}\int_{x_2/2}^{1/2}\frac{dx_1}{x_1}\frac{x_2}{x_1} {\cal
K}_{GLR-MQ-ZRS}\left(\frac{x_2}{x_1},\alpha_s\right)
G^{(2)}(x_1,Q_1^2), \eqno(3.34)$$ where a power suppressed factor
$1/Q^2_1$ has been extracted from the evolution kernel.

    The correlation function $G^{(2)}$ is a generalization of
the gluon distribution beyond the leading twist. It is usually
modeled as the square of the gluon distribution. For example,

$$G^{(2)}(x,Q^2)=\frac{1}{\pi R^2_N}G^2(x,Q^2),\eqno(3.35)$$ where
$R_N$ is the correlation scale of the gluons in the nucleon. The
definition (3.35) is a phenomenological model, which contains an
arbitral normalization constant. However, this constant will be
determined through the value of $R_N$ by using the experimental
data.

    The complete GLR-MQ-ZRS equation includes the contributions
of the two-partons-to-two-partons ($2\rightarrow 2$) amplitude, the
interference amplitude between the one-parton-to-two-partons
($1\rightarrow 2$) amplitude and the three-partons-to-two-partons
($3\rightarrow 2$) amplitude. Where we meet very complicated
calculations about the interference- and corresponding virtual
amplitudes. However, the TOPT-cutting rule shows that the above
mentioned amplitudes correspond to a similar recombination kernel
except the numerical factor and the different kinematic regions [4].

    Another key problem is that we meet various multi-gluon
correlation functions, in which the cut line cuts off the
nonperturbative matrix with different ways. Fortunately, Jaffe has
shown that these correlation functions on the light-cone has the
same form in the DIS processes [13]. The Jaffe-cutting rule was
broadly used in the study of the high twist processes. The TOPT
provides a straightforward explanation about the Jaffe-cutting rule:
since all backward propagators are absorbed into the nonperturbative
correlation functions, the partons correlating two initial gluons
inside the nonperturbative matrix are on mass-shell. Therefore, the
correlation functions with cuts at different places are the same.
Thus, the Jaffe-cutting rule can be included in our TOPT-cutting
rule. Combining the DGLAP dynamics at small $x$, the GLR-MQ-ZRS
equation reads

$$\frac{\partial G(x_B,Q^2)}{\partial\ln Q^2}$$
$$=\frac{\alpha_sN_c}{\pi}\int^1_{x_B}
\frac{dx_1}{x_1}G(x_1,Q^2) +\frac{9\alpha_s^2}{2\pi R^2_N
Q^2}\frac{N_c^2}{N_c^2-1}
\int_{x_B/2}^{1/2}\frac{dx_1}{x_1}G^2(x_1,Q^2)$$
$$-\frac{9\alpha_s^2}{\pi R^2_NQ^2}\frac{N_c^2}{N_c^2-1}
\int_{x_B}^{1/2}\frac{dx_1}{x_1}G^2(x_1,Q^2), \eqno(3.36)$$ where
the contributions of the virtual diagrams are cancelled each other.
The second term on the right hand-side of Eq. (3.36) is the positive
antishadowing part, while the third term is the negative shadowing
part.

    Returning to our new evolution equation. We model the correlation function $F^{(2)}$ as the square of the gluon
distribution as in the leading twist case Eq. (3.35), i.e.,

$$\widetilde{F}^{(2)}(x,\underline{k}_{bc})=\int
d^2\underline{k}\widetilde{R}_F(\underline{k}_{bc},\underline{k})
\widetilde{F}(x,\underline{k}_{bc})\widetilde{F}(x,\underline{k})\equiv\frac{1}{\pi
R^2_N}\widetilde{F}^2(x,\underline{k}_{bc}), \eqno(3.37)$$where we
take the same parameter $R_N$ as in Eq. (3.35) since the relation
(2.28) is irrelevant to $R_N$.  Using the evolution kernel (3.32),
we write

$$\widetilde {F}(x_2,\underline{k}_{b0})=\widetilde {F}(x_1,\underline{k}_{bc})+\Delta \widetilde {F}(x_2,\underline{k}_{b0})$$
$$=\widetilde {F}(x_1,\underline{k}_{bc})
+\frac{9\alpha^2_s}{2\pi^2R^2_N}\frac{N_c^2}{N_c^2-1} \int
d^2\underline{k}_{bc}\int_{x_2/2}^{1/2}\frac{dx_1}{x_1}
\frac{1}{\underline{k}_{bc}^2}\frac{\underline{k}_{bc}^2}{\underline{k}_{b0}^2\underline{k}_{0c}^2}
\widetilde {F}^2(x_1,\underline{k}_{bc}). \eqno(3.38)$$

    Now let us discuss the contributions from the virtual diagrams.
 According to the standard regularization schema, the TOPT-cutting
rule shows that the diagrams in Fig. 7 have a similar evolution
kernel as that in Fig. 6 but with the different kinematical
variables and differ from a simple numerical factor.

    The processes in Figs. 6 and 7 contributes the net positive antishadowing effect.  The negative
shadowing effect is really originated from the interference
processes, two of them are shown in Fig. 8. Here the contributions
from the corresponding virtual processes are also necessary (see
Fig. 9). The TOPT-cutting rule shows that the processes in Figs. 8
and 9 also have a similar evolution kernel.

 \vskip 0.5
truecm \hbox{
\centerline{\epsfig{file=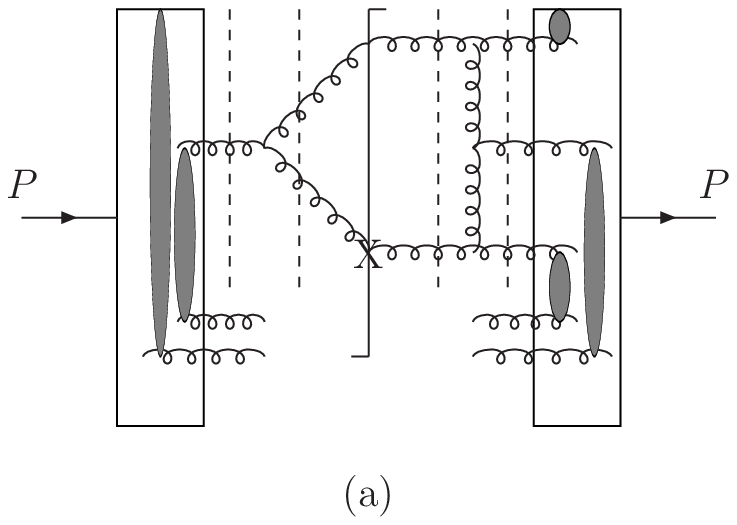,width=5.0cm,clip=}}} \hbox{
\centerline{\epsfig{file=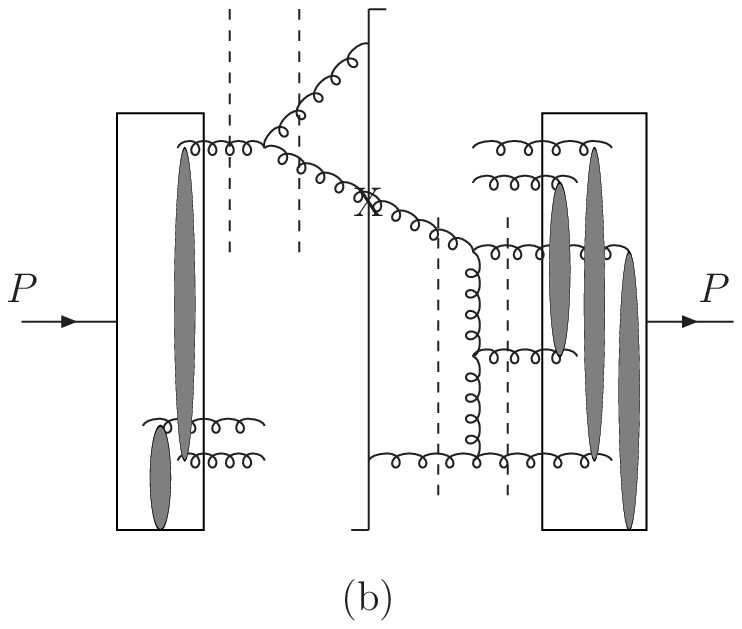,width=5.0cm,clip=}}}
 \vskip 0 truecm

  Fig.8: One of the TOPT-diagrams for the interference
processes, which have the same order as Fig. 6.

 \vskip 1.0 truecm

\hbox{ \centerline{\epsfig{file=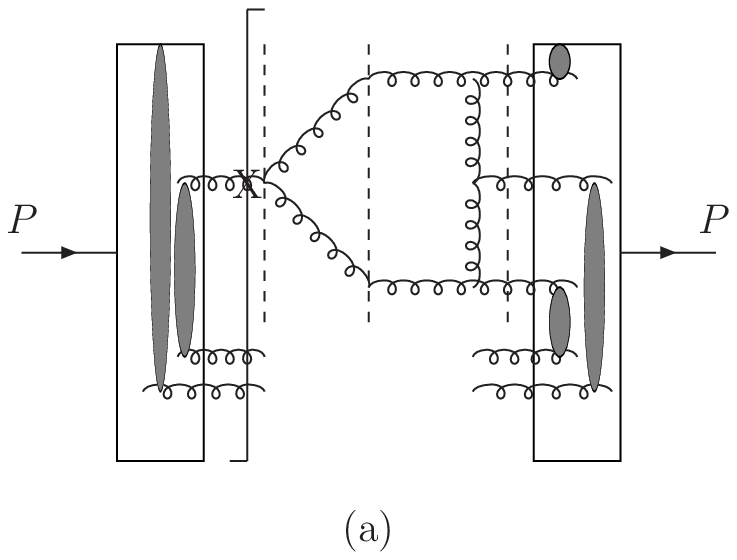,width=5.0cm,clip=}}}
\hbox{ \centerline{\epsfig{file=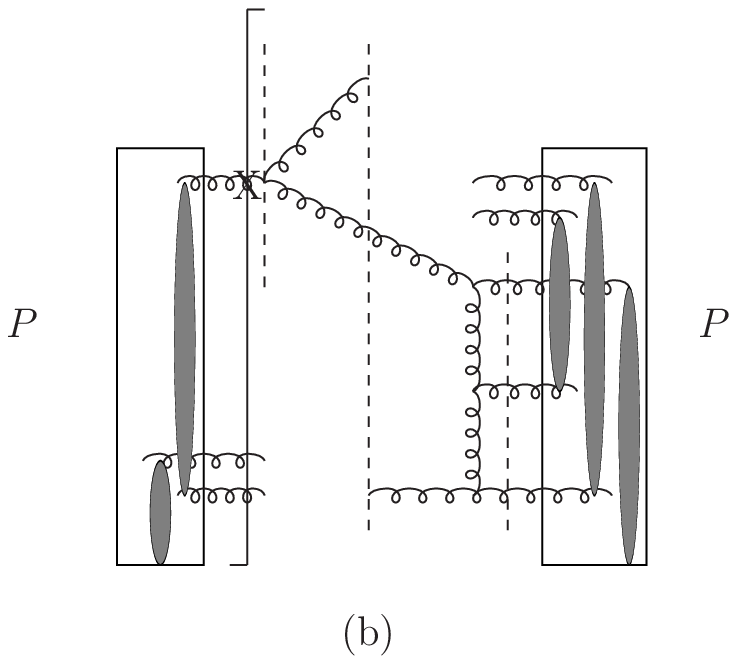,width=5.0cm,clip=}}}

 \vskip 0 truecm

\noindent Fig.9: The Virtual diagrams corresponding to Fig. 8,
they contain a similar evolution kernel but with a different
numerical factor according to the TOPT-cutting rule.

\vskip 1.0 truecm

\hbox{ \centerline{\epsfig{file=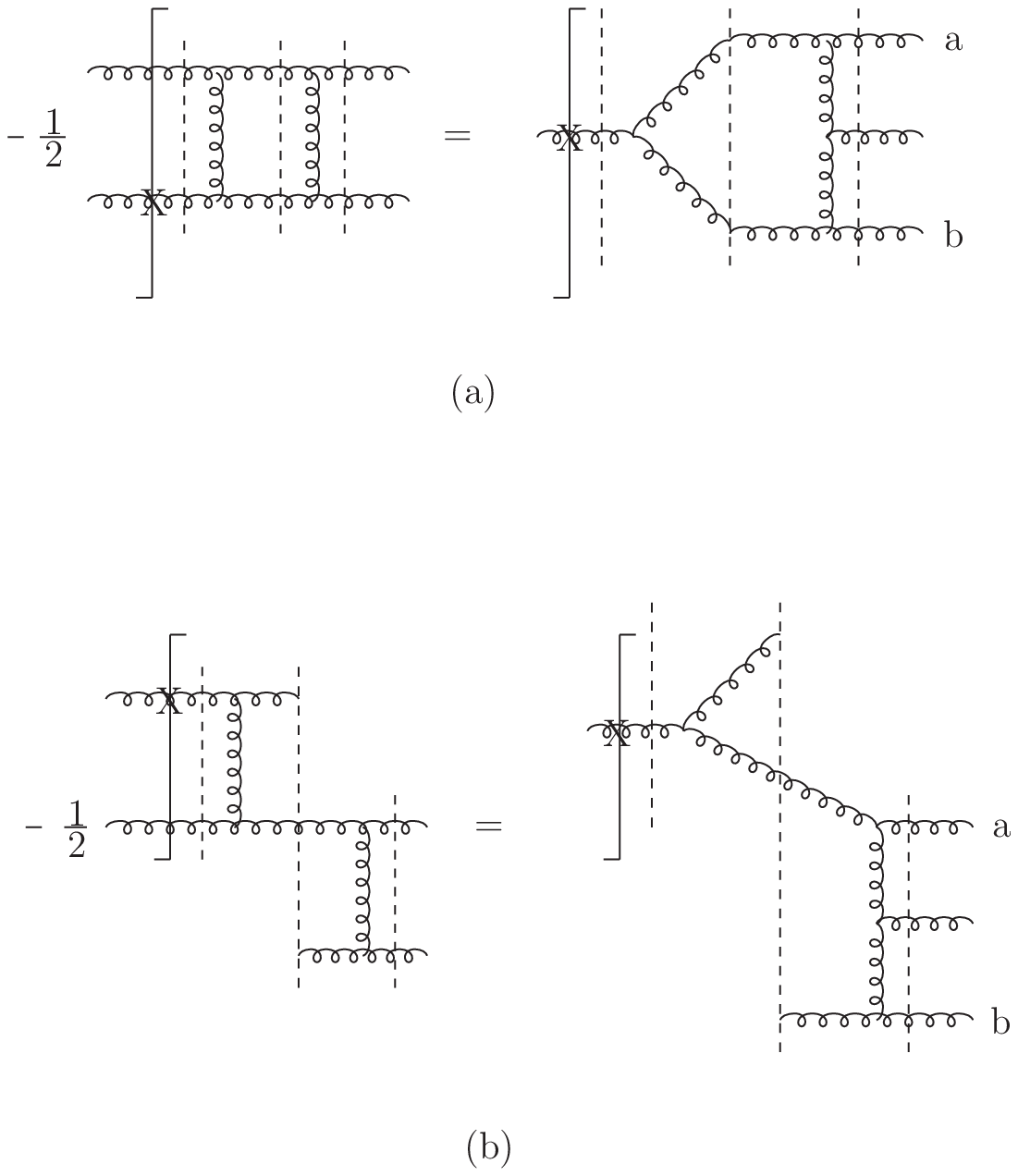,width=5.5cm,clip=}}}
\noindent Fig.10: The TOPT-cutting rule shows a simple relations
among virtual diagrams in Figs. 7 and 9. Thus, all diagrams in
Figs. 6, 7, 8 and 9 contain a similar evolution kernel but with
different numerical factors. \vskip 1.0 truecm

    Up to now we have separately established the relations of the evolution kernels
between the real and virtual diagrams in the 4-partons-to-4 partons
($4\rightarrow 4$) amplitude and the 3-partons-to-5-partons
($3\rightarrow 5$) amplitude, respectively. In the next step we will
show that the relationship between the above mentioned two kinds of
virtual diagrams will link up all the four evolution kernels.
According to Eq. (3.20), the resulting amplitudes are irrelevant to
the transverse momenta of the initial gluons at $x_2\ll x_1$. Thus,
we use the relations shown in Fig. 10, which are derived in the
collinear factorization schema [4] to reveal that the two kinds of
virtual diagrams differ only from a minus sign, which is from an
energy deficit between the two dashed lines in Fig. 9: because both the
momenta $\underline {k}_{b0}$ and $\underline{k}_{0c}$ are indicated
by $\underline{k}$ in the mass-center of the nucleon target, we have

$$\frac{\underline{k}^2}{2x_mP}-\frac{\underline{k}^2}{2x_lP}>0, \eqno(3.39)$$ on
the left-hand side of Fig. 10, where $x_m<x_l$, ($x_m$ and $x_l$ are
the longitudinal momentum fractions in the momenta $m$ and $l$,
respectively); and

$$\frac{\underline{k}^2}{2x_mP}-\frac{\underline{k}^2}{2x_lP}<0, \eqno(3.40)$$ on
the right-hand side of Fig. 10, where $x_m>x_l$.

\vskip -1.0 truecm \hbox{
\centerline{\epsfig{file=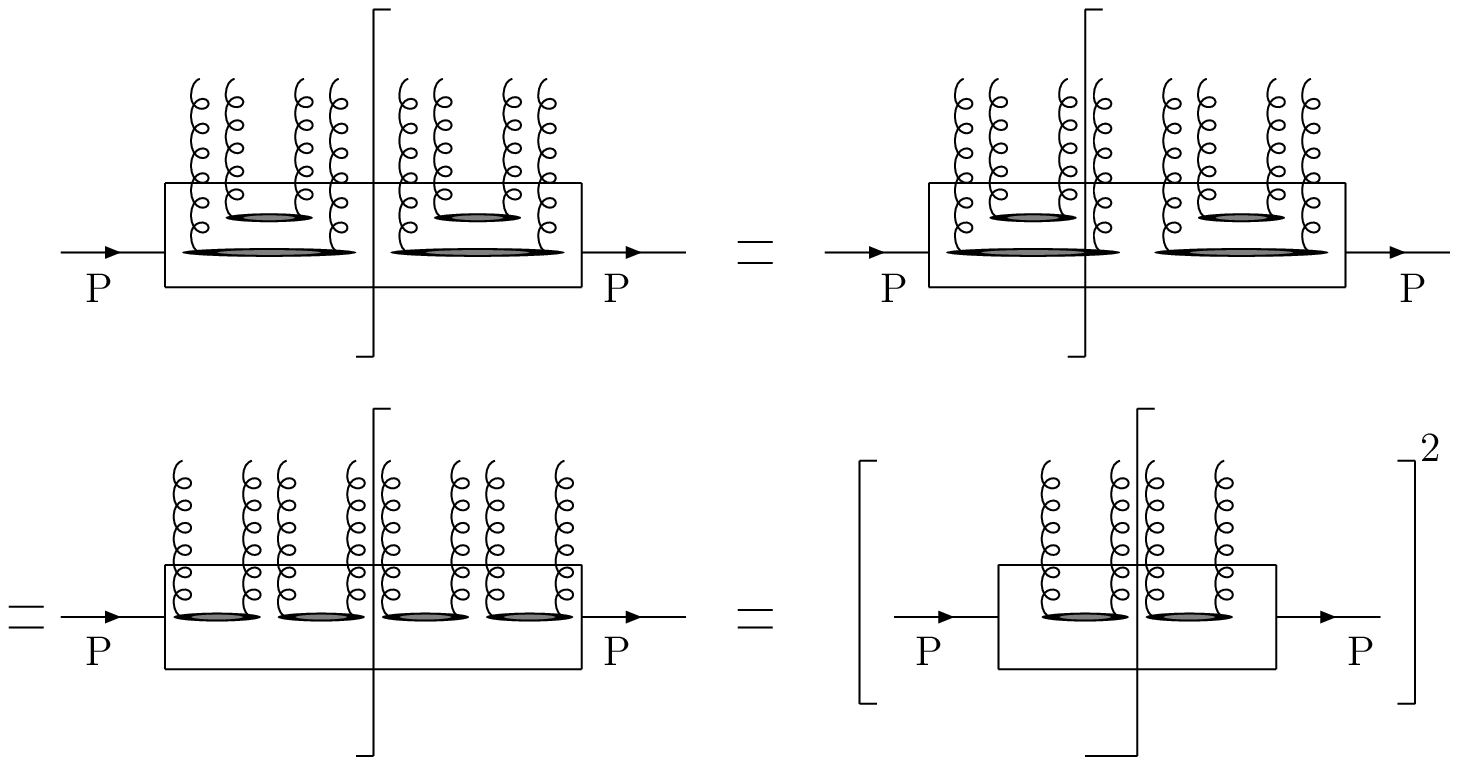,width=12.0cm,clip=}}} \noindent
\vskip -8.5 truecm

Fig.11: The model for the multi-gluons correlating function based
on the TOPT-cutting rules. The propagator inside the cold spot is
forward and on mass-shell at the $W-W$ approximation, while the
correlations to the cold spot from the other part of the
nonperturbative matrix are neglected. Thus, the correlating
function can be cut.

\vskip 1.0 truecm

     In consequence, we finally link up all evolution kernels and
obtain the following equation

$$\widetilde {F}(x_2,\underline{k}_{b0})=\widetilde {F}(x_1,\underline{k}_{bc})+\Delta \widetilde {F}(x_2,\underline{k}_{b0})$$
$$=\widetilde {F}(x_1,\underline{k}_{bc})
+\frac{9\alpha^2_s}{2\pi^2R^2_N}\frac{N^2_c}{N^2_c-1} \int
d^2\underline{k}_{bc}\int_{x_2/2}^{1/2}\frac{dx_1}{x_1}
\frac{1}{\underline{k}_{bc}^2}\frac{\underline{k}_{bc}^2}{\underline{k}_{b0}^2\underline{k}_{0c}^2}
\widetilde {F}^2(x_1,\underline{k}_{bc})$$
$$-\frac{9\alpha^2_s}{4\pi^2R^2_N}\frac{N_c^2}{N_c^2-1}
\int d^2\underline{k}_{bc}\int_{x_2/2}^{1/2}\frac{dx_1}{x_1}
\frac{1}{\underline{k}_{b0}^2}\frac{\underline{k}_{b0}^2}{\underline{k}_{bc}^2\underline{k}_{c0}^2}\widetilde
{F}^2\left(x_1,\underline{k}_{b0}\right)$$
$$-\frac{9\alpha^2_s}{\pi^2R^2_N}\frac{N^2_c}{N^2_c-1} \int
d^2\underline{k}_{bc}\int_{x_2}^{1/2}\frac{dx_1}{x_1}
\frac{1}{\underline{k}_{bc}^2}\frac{\underline{k}_{bc}^2}{\underline{k}_{b0}^2\underline{k}_{0c}^2}
\widetilde {F}^2(x_1,\underline{k}_{bc})$$
$$+\frac{9\alpha^2_s}{2\pi^2R^2_N}\frac{N^2_c}{N^2_c-1} \int d^2\underline{k}_{bc}\int_{x_2}^{1/2}\frac{dx_1}{x_1}
\frac{1}{\underline{k}_{b0}^2}\frac{\underline{k}_{b0}^2}{\underline{k}_{bc}^2\underline{k}_{c0}^2}\widetilde
{F}^2(x,\underline{k}_{b0}) , \eqno(3.41)$$ where we assume that the
Jaffe-cutting rule is still holden in the
$\underline{k}$-factorization scheme (Fig. 11).  The reasons are as
follows: (a) the propagator inside the cold spot is forward and on
mass-shell at the $W-W$ approximation; (b) the correlations to the
cold spot from the other part of the nonperturbative matrix are
neglected in our model Eq. (2.17). Thus, the correlation function
can be cut and we can use the same correlation function in the real,
virtual, and interference processes. From Eq. (3.41) we have

$$-x\frac {\partial \widetilde {F}(x,\underline{k}_{b0})}{\partial x}$$
$$=\frac{9\alpha^2_s}{2\pi^2R^2_N}\frac{N_c^2}{N_c^2-1}
\int d^2\underline{k}_{bc}
\frac{1}{\underline{k}_{bc}^2}\frac{\underline{k}_{bc}^2}{\underline{k}_{b0}^2\underline{k}_{0c}^2}
\widetilde {F}^2\left(\frac{x}{2},\underline{k}_{bc}\right)$$
$$-\frac{9\alpha^2_s}{4\pi^2R^2_N}\frac{N_c^2}{N_c^2-1}\widetilde {F}^2\left(\frac{x}{2},\underline{k}_{b0}\right)
\int d^2\underline{k}_{bc}
\frac{1}{\underline{k}_{b0}^2}\frac{\underline{k}_{b0}^2}{\underline{k}_{bc}^2\underline{k}_{c0}^2}$$
$$-\frac{9\alpha^2_s}{\pi^2R^2_N}\frac{N^2_c}{N^2_c-1} \int
d^2\underline{k}_{bc}
\frac{1}{\underline{k}_{bc}^2}\frac{\underline{k}_{bc}^2}{\underline{k}_{b0}^2\underline{k}_{0c}^2}\widetilde
{F}^2(x,\underline{k}_{bc})$$
$$+\frac{9\alpha^2_s}{2\pi^2R^2_N}\frac{N^2_c}{N^2_c-1} \widetilde
{F}^2(x,\underline{k}_{b0})\int d^2\underline{k}_{bc}
\frac{1}{\underline{k}_{b0}^2}\frac{\underline{k}_{b0}^2}{\underline{k}_{bc}^2\underline{k}_{c0}^2}
 \eqno(3.42)$$

 Similar to Eq. (2.34) we note that

$$\tilde{F}^{(2)}(x,\underline{k}_{bc})\propto
\left[\frac{1}{E_P-2E_ {bc}-E_X}\right]^2\sim
\frac{1}{\underline{k}^4_{bc}}. \eqno(3.43)$$

    we redefine

$$F^{(2)}(x,\underline{k})
=\int d^2\underline{k}'R_F(\underline{k},\underline{k}')
F(x,\underline{k})F(x,\underline{k}')$$
$$\equiv \frac{\underline{k}^4}{\hat{\underline{k}}^4}
\tilde{F}^{(2)}(x,\underline{k}),   \eqno(3.44)$$ where
$R_F=\widetilde{R}_F\hat{\underline{k}}^4/\underline{k}^4$.
Submitting this equation with Eq. (2.34) to Eq. (3.42), the result
is

$$-x\frac {\partial F(x,\underline{k}_{b0})}{\partial x}$$
$$=\frac{9\alpha^2_s}{2\pi^2R^2_N}\frac{N_c^2}{N_c^2-1}
\int d^2\underline{k}_{bc}
\frac{1}{\underline{k}_{bc}^2}\frac{\underline{k}_{b0}^2}{\underline{k}_{bc}^2\underline{k}_{c0}^2}
F^2\left(\frac{x}{2},\underline{k}_{bc}\right)$$
$$-\frac{9\alpha^2_s}{4\pi^2R^2_N}\frac{N_c^2}{N_c^2-1}F^2\left(\frac{x}{2},\underline{k}_{b0}\right)
\int d^2\underline{k}_{bc}
\frac{1}{\underline{k}_{b0}^2}\frac{\underline{k}_{b0}^2}{\underline{k}_{bc}^2\underline{k}_{c0}^2}$$
$$-\frac{9\alpha^2_s}{\pi^2R^2_N}\frac{N_c^2}{N_c^2-1} \int
d^2\underline{k}_{bc}
\frac{1}{\underline{k}_{bc}^2}\frac{\underline{k}_{b0}^2}{\underline{k}_{bc}^2\underline{k}_{c0}^2}
F^2(x,\underline{k}_{bc})$$
$$+\frac{9\alpha^2_s}{2\pi^2R^2_N}\frac{N_c^2}{N_c^2-1}F^2(x,\underline{k}_{b0}) \int
d^2\underline{k}_{bc}
\frac{1}{\underline{k}_{b0}^2}\frac{\underline{k}_{b0}^2}{\underline{k}_{bc}^2\underline{k}_{c0}^2}.
\eqno(3.45)$$

    Combining it with the linear BFKL equation, we finally obtain a complete evolution equation at small
$x$

$$-x\frac {\partial F(x,\underline{k}_{b0})}{\partial x}$$
$$=\frac{\alpha_{s}N_c}{2\pi^2}\int d^2
\underline{k}_{bc}
\frac{\underline{k}_{b0}^2}{\underline{k}_{bc}^2\underline{k}_{c0}^2}
2F(x,\underline{k}_{bc})-\frac{\alpha_{s}N_c}{2\pi^2}
F(x,\underline{k}_{b0})\int d^2 \underline{k}_{bc}
\frac{\underline{k}_{b0}^2}{\underline{k}_{bc}^2\underline{k}_{c0}^2}$$
$$+\frac{9\alpha^2_s}{2\pi^2R^2_N}\frac{N_c^2}{N_c^2-1}
\int d^2\underline{k}_{bc}
\frac{1}{\underline{k}_{bc}^2}\frac{\underline{k}_{b0}^2}{\underline{k}_{bc}^2\underline{k}_{c0}^2}F^2\left(\frac{x}{2},\underline{k}_{bc}\right)
-\frac{9\alpha^2_s}{4\pi^2R^2_N}\frac{N_c^2}{N_c^2-1}F^2\left(\frac{x}{2},\underline{k}_{b0}\right)
\int d^2\underline{k}_{bc}
\frac{1}{\underline{k}_{b0}^2}\frac{\underline{k}_{b0}^2}{\underline{k}_{bc}^2\underline{k}_{c0}^2}$$
$$-\frac{9\alpha^2_s}{\pi^2R^2_N}\frac{N_c^2}{N_c^2-1} \int
d^2\underline{k}_{bc}
\frac{1}{\underline{k}_{bc}^2}\frac{\underline{k}_{b0}^2}{\underline{k}_{bc}^2\underline{k}_{c0}^2}
F^2(x,\underline{k}_{bc})
+\frac{9\alpha^2_s}{2\pi^2R^2_N}\frac{N_c^2}{N_c^2-1}F^2(x,\underline{k}_{b0})
\int d^2\underline{k}_{bc}
\frac{1}{\underline{k}_{b0}^2}\frac{\underline{k}_{b0}^2}{\underline{k}_{bc}^2\underline{k}_{c0}^2}.
\eqno(3.46)$$  Comparing with the GLR-MQ-ZRS equation (3.36), the
contributions of the virtual diagrams can't be canceled in Eq.
(3.46) and they are necessary for IR safety.

\newpage
\begin{center}
\section{Unity of the QCD evolution equations}
\end{center}

     It is a surprise that Eq. (3.46) can be "directly" written by using
an analogy with the DGLAP, BFKL and GLR-MQ-ZRS equations. For this
sake, we summarize the four evolution equations at small $x$ as follows.
The DGLAP equation (2.37)

\vskip -2.0 truecm \hbox{
\centerline{\epsfig{file=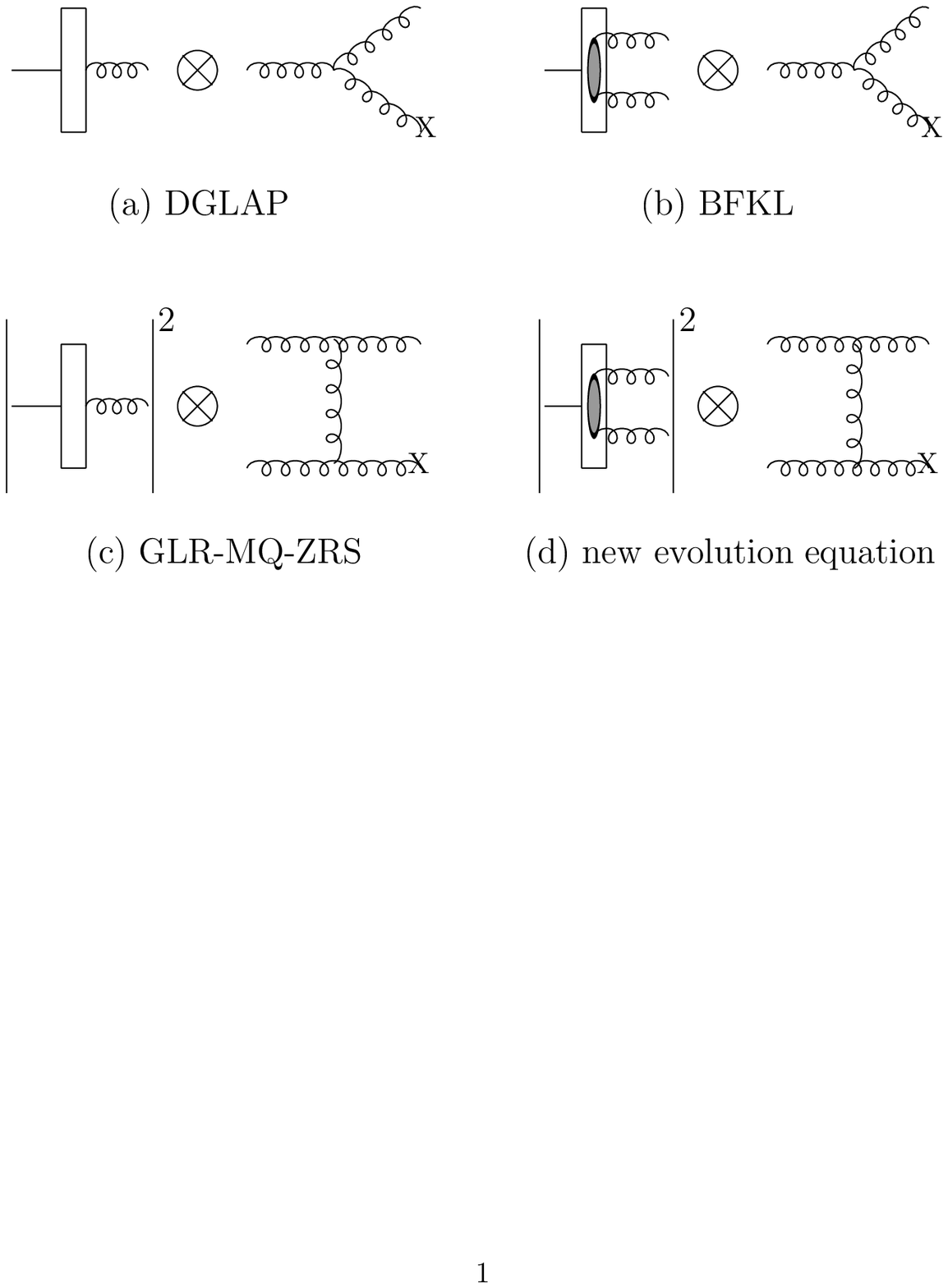,width=12.0cm,clip=}}} \noindent
\vskip -8.0 truecm

Fig.12: The elemental amplitudes for the four evolution equations
based on Fig. 1. \vskip 1.0 truecm

$$Q^2\frac{\partial G(x,Q^2)}{\partial Q^2}=\frac{\alpha_sN_c}{\pi}\int_x^1\frac{dx_1}{x_1}
G(x_1,Q^2), $$ or

$$\Delta G(x,Q^2)=\frac{\alpha_s N_c}{\pi}\int_x^1\frac{dx_1}{x_1}\int^{Q^2}\frac{d\underline{k}^2}{\underline{k}^2}
G(x_1,\underline{k}^2) \eqno(4.1)$$ (see Fig. 12a); The BFKL
equation (2.35)

$$-x\frac {\partial F(x,\underline{k}_{a0})}{\partial x}$$
$$=\frac{\alpha_{s}N_c}{\pi}\int \frac{d^2
\underline{k}_{ab}}{\pi}
\frac{\underline{k}_{a0}^2}{\underline{k}_{ab}^2\underline{k}_{0b}^2}
\left[F(x,\underline{k}_{ab})-\frac{1}{2}F(x,\underline{k}_{a0})\right],$$
or its real part

$$\Delta F(x,\underline{k}_{a0})=\frac{\alpha_{s}N_c}{\pi}
\int_x^1\frac{dx_1}{x_1}\int \frac{d^2 \underline{k}_{ab}}{\pi}
\frac{\underline{k}_{a0}^2}{\underline{k}_{ab}^2\underline{k}_{0b}^2}
F(x_1,\underline{k}_{ab}). \eqno(4.2)$$ (see Fig. 12b); The
GLR-MQ-ZRS equation (3.36)

$$\frac{\partial G(x,Q^2)}{\partial\ln Q^2}$$
$$=\frac{\alpha_sN_c}{\pi}\int^1_x
\frac{dx_1}{x_1}G(x_1,Q^2)$$
$$+\frac{9\alpha_s^2}{2\pi R^2_N
Q^2}\frac{N_c^2}{N_c^2-1}
\int_{x/2}^{1/2}\frac{dx_1}{x_1}G^2(x_1,Q^2)$$
$$-\frac{9\alpha_s^2}{\pi R^2_NQ^2}\frac{N_c^2}{N_c^2-1}
\int_x^{1/2}\frac{dx_1}{x_1}G^2(x_1,Q^2), $$

or

$$\Delta G(x,Q^2)$$
$$=\frac{\alpha_sN_c}{\pi}\int^1_x
\frac{dx_1}{x_1}\int^{Q^2}\frac{d\underline{k}^2}{\underline{k}^2}G(x_1,\underline{k}^2)$$
$$+\frac{9\alpha_s^2}{2\pi R^2_N}\frac{N_c^2}{N_c^2-1}
\int_{x/2}^{1/2}\frac{dx_1}{x_1}\int^{Q^2}\frac{d\underline{k}^2}{\underline{k}^2}\frac{1}{\underline{k}^2}
G^2(x_1,\underline{k}^2)$$
$$-\frac{9\alpha_s^2}{\pi R^2_N}\frac{N_c^2}{N_c^2-1}
\int_x^{1/2}\frac{dx_1}{x_1}\int^{Q^2}\frac{d\underline{k}^2}{\underline{k}^2}\frac{1}{\underline{k}^2}
G^2(x_1,\underline{k}^2) \eqno(4.3)$$ (see Fig. 12c);

    The equation (3.46)

$$-x\frac {\partial F(x,\underline{k}_{b0})}{\partial x}$$
$$=\frac{\alpha_{s}N_c}{\pi}\int d^2
\frac{\underline{k}_{bc}}{\pi}
\frac{\underline{k}_{b0}^2}{\underline{k}_{bc}^2\underline{k}_{c0}^2}
\left[F(x,\underline{k}_{bc})-\frac{1}{2}F(x,\underline{k}_{b0})\right]$$
$$+\frac{9\alpha^2_s}{2\pi R^2_N}\frac{N_c^2}{N_c^2-1}
\int \frac{d^2\underline{k}_{bc}}{\pi}
\frac{\underline{k}_{b0}^2}{\underline{k}_{bc}^2\underline{k}_{c0}^2}\left[\frac{1}{\underline{k}_{bc}^2}
F^2\left(\frac{x}{2},\underline{k}_{bc}\right)-\frac{1}{2\underline{k}_{b0}^2}
F^2\left(\frac{x}{2},\underline{k}_{b0}\right)\right ]$$
$$-\frac{9\alpha^2_s}{\pi R^2_N}\frac{N_c^2}{N_c^2-1} \int
\frac{d^2\underline{k}_{bc}}{\pi}
\frac{\underline{k}_{b0}^2}{\underline{k}_{bc}^2\underline{k}_{c0}^2}
\left [\frac{1}{\underline{k}_{bc}^2}F^2(x,\underline{k}_{bc})
-\frac{1}{2\underline{k}_{b0}^2}F^2(x,\underline{k}_{b0})\right].
$$ or its real part

$$\Delta F(x,\underline{k}_{b0})$$
$$=\frac{\alpha_{s}N_c}{\pi}\int_x^1\frac{dx_1}{x_1}\int \frac{d^2
\underline{k}_{bc}}{\pi}
\frac{\underline{k}_{b0}^2}{\underline{k}_{bc}^2\underline{k}_{c0}^2}
F(x_1,\underline{k}_{bc})$$
$$+\frac{9\alpha^2_s}{2\pi R^2_N}\frac{N_c^2}{N_c^2-1}\int_{x/2}^{1/2}\frac{dx_1}{x_1} \int
\frac{d^2\underline{k}_{bc}}{\pi}
\frac{\underline{k}_{b0}^2}{\underline{k}_{bc}^2\underline{k}_{c0}^2}
\frac{1}{\underline{k}_{bc}^2}F^2(x_1,\underline{k}_{bc})$$
$$-\frac{9\alpha^2_s}{\pi R^2_N}\frac{N_c^2}{N_c^2-1}\int_x^{1/2}\frac{dx_1}{x_1} \int
\frac{d^2\underline{k}_{bc}}{\pi}
\frac{\underline{k}_{b0}^2}{\underline{k}_{bc}^2\underline{k}_{c0}^2}
\frac{1}{\underline{k}_{bc}^2}F^2(x_1,\underline{k}_{bc})
\eqno(4.4)$$ (see Fig. 12d).

    One can find the following interesting relations among these
equations: The DGLAP and BFKL equations have the same evolution
dynamics (i.e., the gluon splitting), where we have the following
analogy between the real parts of Eqs. (4.1) and (4.2):

$$\frac{d\underline{k}^2}{\underline{k}^2}\leftrightarrow
\frac{d^2\underline{k}_{ab}}{\pi}\frac{\underline{k}^2_{a0}}{\underline{k}^2_{ab}\underline{k}^2_{0b}},\eqno(4.5)$$

$$G(x,\underline{k}^2)\leftrightarrow
F(x,\underline{k}_{ab}). \eqno(4.6)$$

    The nonlinear parts of the GLR-MQ-ZRS and Eq. (3.46) also
have the same evolution dynamics (i.e., the gluon recombination),
they have similar relationships like Eqs. (4.5) and (4.6):

$$\frac{d\underline{k}^2}{\underline{k}^2}\leftrightarrow
\frac{d^2\underline{k}_{bc}}{\pi}\frac{\underline{k}^2_{b0}}{\underline{k}^2_{bc}\underline{k}^2_{c0}},\eqno(4.7)$$

$$G(x,\underline{k}^2)\leftrightarrow F(x,\underline{k}_{bc}),\eqno(4.8)$$
and an extra relation for the power suppression factor

$$\frac{1}{\underline{k}^2}\leftrightarrow\frac{1}{\underline{k}^2_{bc}}~or~\frac{1}{\underline{k}^2_{b0}} .
\eqno(4.9)$$ Thus, we can directly write the real part (4.4) of Eq.
(3.46) following the DGLAP, BFKL and GLR-MQ-ZRS equations.

    A complete evolution equation includes the contributions from all possible
cut diagrams at a given order, including the virtual diagrams for
the regularization of the theory. The resulting evolution equations
(2.37), (2.35), (3.36) and (3.46) have the following structure:

$$DGLAP: real~ part-virtual ~part^{a},\eqno(4.10)$$

$$BFKL:  real~ part-virtual~ part,\eqno(4.11)$$

$$GLR-MQ-ZRS (nonlinear~ part): real ~part-virtual~
    part^{b}$$$$-real~interferance~part+virtual~
                                 interferance~part^{c},\eqno(4.12)$$

$$Eq. (3.46)(nonlinear~ part): real ~part-virtual~part$$
$$-real~ interferance~ part+virtual~interferance~
part,\eqno(4.13)$$where the contributions of the virtual cut
diagrams play an important role, although (a) is neglected at small
$x$ [1], (b) and (c) are cancelled each other after the relations
established among the different cut diagrams [4]. According to the
TOPT cutting rule, the four nonlinear terms in Eq. (4.13) share a
common evolution kernel, and they differ only by a numerical factor
($\pm 1$ or $\pm 1/2$) and the integration range. Thus, we can write
the complete Eq. (3.46) based on Eq. (4.4).

    The BK equation [5] is generally considered as a typical
nonlinear correction to the BFKL equation at the $LL(1/x)$
approximation. We discuss the relation of Eq. (3.46) with the BK
equation. The BK equation is usually written by using the scattering
amplitude $N(\underline{x},x)$ in the transverse coordinator space

$$-x\frac{\partial N(\underline{x}_{b0},x)}{\partial x}$$
$$=\frac{\alpha_{s}N_c}{2\pi^2}\int d^2 \underline{x}_{c}
\frac{\underline{x}_{b0}^2}{\underline{x}_{bc}^2\underline{x}_{c0}^2}\left[N(\underline{x}_{bc},x)+N(\underline{x}_{c0},x)-N(\underline{x}_{b0},x)\right.$$
$$-N(\underline{x}_{bc},x)N(\underline{x}_{c0},x)].\eqno(4.14)$$
The nonlinear evolution kernel in the BK equation is regularized by
the connecting amplitude
$N(\underline{x}_{bc},x)N(\underline{x}_{c0},x)$ rather than using
the virtual diagrams. Using

$$N(x,\underline{k})=\int
\frac{d^2\underline{x}}{2\pi}\exp(-i\underline{k}\cdot\underline{x})
\frac{N(\underline{x},x)}{\underline{x}^2},\eqno(4.15)$$ and the
definition

$$N(x,\underline{k})\equiv
\frac{27\alpha_s}{16\underline{k}^2R^2_N}F(x,\underline{k})
,\eqno(4.16)$$ one can obtain the BK equation in the momentum space

$$-x\frac{\partial F(x,\underline{k}_{b0})}{\partial x}$$
$$=\frac{\alpha_{s}N_c}{2\pi^2}\int d^2
\underline{k}_{bc}
\frac{\underline{k}_{b0}^2}{\underline{k}_{bc}^2\underline{k}_{c0}^2}
2F(x,\underline{k}_{bc})-\frac{\alpha_{s}N_c}{2\pi^2}
F(x,\underline{k}_{b0})\int d^2 \underline{k}_{bc}
\frac{\underline{k}_{b0}^2}{\underline{k}_{bc}^2\underline{k}_{c0}^2}$$
$$-\frac{9\alpha_s^2}{2\pi R^2_N}\frac{N_c^2}{N_c^2-1}\frac{1}{\underline{k}^2_{b0}}
F^2(x,\underline{k}_{b0}). \eqno(4.17)$$ Since the measured unintegrated gluon distribution $F(x,\underline{k}^2)$
is irrelevant to the azimuthal angle $\phi$ (see Eq. 2.38), after azimuthal integration we have

$$-x\frac{\partial F(x,\underline{k}^2)}{\partial x}$$
$$=\frac{3\alpha_{s}\underline{k}^2}{\pi}\int_{\underline{k}^2_0}^{\infty} \frac{d \underline{k
}'^2}{\underline{k}'^2}\left\{\frac{F(x,\underline{k}'^2)-F(x,\underline{k}^2)}
{\vert
\underline{k}'^2-\underline{k}^2\vert}+\frac{F(x,\underline{k}^2)}{\sqrt{\underline{k}^4+4\underline{k}'^4}}\right\}
-\frac{81}{16}\frac{\alpha_s^2}{\pi
R^2_N}\frac{1}{\underline{k}^2}F^2(x,\underline{k}^2). \eqno(4.18)$$
The similar form of the BK equation in the momentum configuration
was used by other authors [14] with a different definition (4.16).
We call Eq. (4.18) as the BK-like equation.

    Now we derive Eq. (4.17) but from Eq. (3.46). For this sake, we
remove the contributions of Figs. 6c and 6d in the derivation of Eq.
(3.46) according to Fig. 12d. Thus, Eq. (4.4) reduces to

$$\Delta F(x,\underline{k}_{b0})$$
$$=\frac{\alpha_{s}N_c}{\pi}\int_x^1\frac{dx_1}{x_1}\int \frac{d^2
\underline{k}_{bc}}{\pi}
\frac{\underline{k}_{b0}^2}{\underline{k}_{bc}^2\underline{k}_{c0}^2}
F(x_1,\underline{k}_{bc})$$
$$+\frac{9\alpha^2_s}{2\pi R^2_N}\frac{N_c^2}{N_c^2-1}\int_{x/2}^{1/2}\frac{dx_1}{x_1}
\int^{\underline{k}^2_{b0}}_{\underline{k}^2_{min}}\frac{d\underline{k}^2_{bc}}{\underline{k}^2_{bc}}\frac{1}{\underline{k}^2_{bc}}
F^2(x_1,\underline{k}_{bc})$$
$$-\frac{9\alpha^2_s}{\pi
R^2_N}\frac{N_c^2}{N_c^2-1}\int_x^{1/2}\frac{dx_1}{x_1}
\int^{\underline{k}^2_{b0}}_{\underline{k}^2_{min}}\frac{d\underline{k}^2_{bc}}{\underline{k}^2_{bc}}\frac{1}{\underline{k}^2_{bc}}
F^2(x_1,\underline{k}_{bc}), \eqno(4.19)$$where we use Eq. (3.33),
i.e.,

$$\int \frac{d^2\underline{k}_{bc}}{\pi}
\frac{\underline{k}_{b0}^2}{\underline{k}_{bc}^2\underline{k}_{c0}^2}
\frac{1}{\underline{k}_{bc}^2}\rightarrow
\int^{\underline{k}^2_{b0}}_{\underline{k}^2_{min}}\frac{d\underline{k}^2_{bc}}{\underline{k}^2_{bc}}\frac{1}{\underline{k}^2_{bc}}.
\eqno(4.20)$$

    The nonlinear evolution kernel in Eq. (4.19) is essentially
the GLR-MQ-ZRS-kernel [12] and it collects only the
$\underline{k}^2$-ordered corrections. That is,
$\underline{k}^2_{bc}$ are ordered in
$[\underline{k}^2_{min},\underline{k}^2_{b0}]$. As an approximation,
we only keep the last step evolution, i.e., we set
$\underline{k}^2_{bc}=\underline{k}^2_{b0}$ and call it as the one
step evolution approximation. Insert the dimensionless function
$\delta(1-\underline{k}^2_{b0}/\underline{k}^2_{bc})$ into Eq.
(4.19),

$$\Delta F(x,\underline{k}_{b0})$$
$$=\frac{\alpha_{s}N_c}{\pi}\int_x^1\frac{dx_1}{x_1}\int \frac{d^2
\underline{k}_{bc}}{\pi}
\frac{\underline{k}_{b0}^2}{\underline{k}_{bc}^2\underline{k}_{c0}^2}
F(x_1,\underline{k}_{bc})$$ $$+\frac{9\alpha^2_s}{2\pi
R^2_N}\frac{N_c^2}{N_c^2-1}\int_{x/2}^{1/2}\frac{dx_1}{x_1}
\frac{1}{\underline{k}^2_{b0}} F^2(x_1,\underline{k}_{b0})
-\frac{9\alpha^2_s}{\pi
R^2_N}\frac{N_c^2}{N_c^2-1}\int_x^{1/2}\frac{dx_1}{x_1}
\frac{1}{\underline{k}^2_{b0}} F^2(x_1,\underline{k}_{b0}),
\eqno(4.21)$$ which leads to the BK-like equation (4.17)

$$-x\frac{\partial F(x,\underline{k}_{b0})}{\partial x}$$
$$=\frac{\alpha_{s}N_c}{2\pi^2}\int d^2
\underline{k}_{bc}
\frac{\underline{k}_{b0}^2}{\underline{k}_{bc}^2\underline{k}_{c0}^2}
2F(x,\underline{k}_{bc})-\frac{\alpha_{s}N_c}{2\pi^2}
F(x,\underline{k}_{b0})\int d^2 \underline{k}_{bc}
\frac{\underline{k}_{b0}^2}{\underline{k}_{bc}^2\underline{k}_{c0}^2}$$
$$+\frac{9\alpha_s^2}{2\pi R^2_N}\frac{N_c^2}{N_c^2-1}\frac{1}{\underline{k}^2_{b0}}
F^2(\frac{x}{2},\underline{k}_{b0})-\frac{9\alpha_s^2}{\pi
R^2_N}\frac{N_c^2}{N_c^2-1}\frac{1}{\underline{k}^2_{b0}}
F^2(x,\underline{k}_{b0})$$
$$\simeq\frac{\alpha_{s}N_c}{2\pi^2}\int d^2
\underline{k}_{bc}
\frac{\underline{k}_{b0}^2}{\underline{k}_{bc}^2\underline{k}_{c0}^2}
2F(x,\underline{k}_{bc})-\frac{\alpha_{s}N_c}{2\pi^2}
F(x,\underline{k}_{b0})\int d^2 \underline{k}_{bc}
\frac{\underline{k}_{b0}^2}{\underline{k}_{bc}^2\underline{k}_{c0}^2}$$
$$-\frac{9\alpha_s^2}{2\pi R^2_N}\frac{N_c^2}{N_c^2-1}\frac{1}{\underline{k}^2_{b0}}
F^2(x,\underline{k}_{b0}), \eqno(4.22)$$where we take
$F^2(x/2,\underline{k}_{bc})\simeq F^2(x,\underline{k}_{bc})$ near
the saturation range. The above derivation of the BK-like equation
indicates that the BK-like equation is a part of Eq. (3.46), where
the contributions from some of the interference sub-processes in Figs.
6c and 6d are removed.

    Therefore, we can regard Eq. (3.46) as a natural expansion of the DGLAP,
BFKL, GLR-MQ-ZRS and BK equations.

\newpage
\begin{center}
\section{Chaos in the new evolution equation}
\end{center}

We will focus on the behavior of the solutions near the saturation
range, where we estimate that $F(x/2,\underline{k}^2)\simeq
F(x,\underline{k}^2)$ in Eq. (3.46) due to the strong shadowing
effect. Thus, Eq. (3.46) reduces to

$$-x\frac {\partial F(x,\underline{k}_{b0})}{\partial x}$$
$$=\frac{\alpha_{s}N_c}{2\pi^2}\int d^2
\underline{k}_{bc}
\frac{\underline{k}_{b0}^2}{\underline{k}_{bc}^2\underline{k}_{c0}^2}
2F(x,\underline{k}_{bc})-\frac{\alpha_{s}N_c}{2\pi^2}
F(x,\underline{k}_{b0})\int d^2 \underline{k}_{bc}
\frac{\underline{k}_{b0}^2}{\underline{k}_{bc}^2\underline{k}_{c0}^2}$$
$$-\frac{9\alpha^2_s}{2\pi^2R^2_N}\frac{N_c^2}{N_c^2-1}
\int d^2\underline{k}_{bc}
\frac{1}{\underline{k}_{bc}^2}\frac{\underline{k}_{b0}^2}
{\underline{k}_{bc}^2\underline{k}_{c0}^2}F^2\left(x,\underline{k}_{bc}\right)
+\frac{9\alpha^2_s}{4\pi^2R^2_N}\frac{N_c^2}{N_c^2-1}F^2\left(x,\underline{k}_{b0}\right)
\int d^2\underline{k}_{bc}
\frac{1}{\underline{k}_{b0}^2}\frac{\underline{k}_{b0}^2}{\underline{k}_{bc}^2\underline{k}_{c0}^2}.
\eqno(5.1)$$

    After azimuthal integration we have

$$-x\frac{\partial F(x,\underline{k}^2)}{\partial x}$$
$$=\frac{3\alpha_{s}\underline{k}^2}{\pi}\int_{\underline{k}^2_0}^{\infty} \frac{d \underline{k
}'^2}{\underline{k}'^2}\left\{\frac{F(x,\underline{k}'^2)-F(x,\underline{k}^2)}
{\vert
\underline{k}'^2-\underline{k}^2\vert}+\frac{F(x,\underline{k}^2)}{\sqrt{\underline{k}^4+4\underline{k}'^4}}\right\}$$
$$-\frac{81}{16}\frac{\alpha_s^2}{\pi R^2_N}\int_{\underline{k}^2_0}^{\infty} \frac{d \underline{k
}'^2}{\underline{k}'^2}\left\{\frac{\underline{k}^2F^2(x,\underline{k}'^2)-\underline{k}'^2F^2(x,\underline{k}^2)}
{\underline{k}'^2\vert
\underline{k}'^2-\underline{k}^2\vert}+\frac{F^2(x,\underline{k}^2)}{\sqrt{\underline{k}^4+4\underline{k}'^4}}\right\}
\eqno(5.2)$$

    The solutions of Eq.(5.2) depend on the strength of the
nonlinear terms, which include the model-dependent assumptions in
Eqs. (3.35), (3.37) and a free parameter $R_N$. To reduce the
uncertainty, the value of $R_N=4GeV^{-1}$ with the assumption (3.37)
is independently fixed by fitting the available experimental data
about the proton structure function using the GLR-MQ-ZRS equation in
Ref. [15].

     The solutions of Eq. (5.2)
need the knowledge of the gluon distribution with all
$\underline{k}^2$ at a starting $x_0$. A major difficulty is the
treatment of the infrared region,
$\underline{k}^2<\underline{k}^2_0$ $(\underline{k}^2_0\sim
1~GeV^2)$. The BFKL evolution leads to diffusion of the starting
$\underline{k}$-distribution both to larger and to smaller values of
$\underline{k}$. However, the perturbative BFKL-growth of
$F(x,\underline{k}^2)$ toward smaller $\underline{k}^2$ is not
expected to be valid when the gluon momenta enter the
nonperturbative region. The common feature of nonperturbative
modifications of the infrared region is that the solution
$F(x,\underline{k}^2)$ vanishes as $\underline{k}^2\rightarrow 0$.
The reasons, for example, are the requirement of gauge invariance
[16], the colour neutrality of the probed proton [17], and the
absence of the valence gluons in a static proton [15]. Therefore,
the increasing distribution $F(x,\underline{k}^2)$ should be
saturated at $\underline{k}^2<Q^2_s(x)$, $Q_s(x)$ is called as the
saturation scale.

\vskip -3.0 truecm \hbox{
\centerline{\epsfig{file=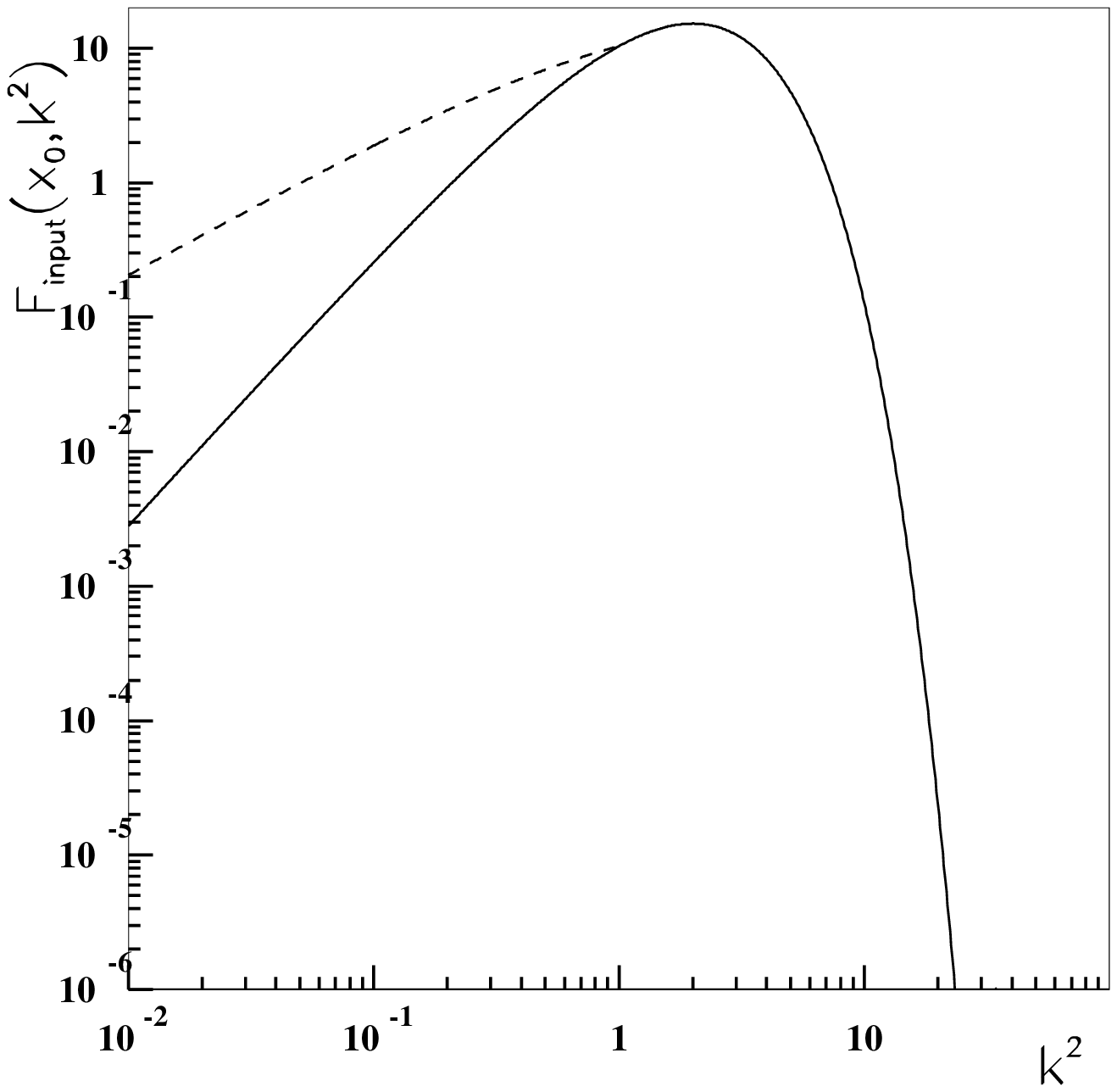,width=12.0cm,clip=}}} \noindent
\vskip -4.5 truecm

Fig.13: The input distributions at $x_0=0.4\times 10^{-4}$; the
solid curve: based on the GBW model Eq. (5.4) and the dashed
curve: based on the saturation model (6.2).

 \vskip 1.0 truecm

For example, with the color-dipole approach Golec-Biernat and Wusthoff (GBW) [18] used
the inclusive and diffractive scattering data and obtained

$${\mathcal{F}}_{GBW}(x,\underline{k}^2)=\frac{3\sigma_0}{4\pi^2\overline{\alpha}_s}R^2_0(x)\underline{k}^2
\exp(-R^2_0(x)\underline{k}^2), \eqno(5.3)$$where
$\sigma_0=29.12mb$, $x_0=0.4\times 10^{-4}$, $\lambda=0.277$,
$R_0(x)=(x/x_0)^{\lambda/2}/Q_s$ and $Q_s=1GeV$. Note
${\mathcal{F}}\equiv F/\underline{k}^2$ in Eq. (2.38). The parameter
$\overline{\alpha}_s$ is fixed as $\overline{\alpha}_s=0.2$. The GBW
model gives a description of $F$ near the saturation scale, although
it lacks the QCD evolution. We draw
$F_{input}(x_0,\underline{k}^2)=\underline{k}^2{\mathcal{F}}_{GBW}(x_0,\underline{k}^2)$
in Fig. 13 (the solid curve). In the calculations we divide the
evolution region into two parts: region(A) 0 to $Q^2_s$ and
region(B) $Q^2_s$ to $\infty$. In region(B) the QCD evolution
equation is taken to evolute and in region(A) the nonperturbative
part of $F(x,k^2)$ is identified as

$$F(x,\underline{k}^2)=C\underline{k}^2{\mathcal{F}}_{GBW}(x,\underline{k}^2),
~~at~x\leq x_0,~\underline{k}^2\leq Q^2_s,\eqno(5.4)$$ where the
parameter $C$ keeps the connection between two parts.

    The Runge-Kutta method is used to compute Eq. (5.2).
Note that $F(x,\underline{k}^2)=0$ if $F(x,\underline{k}^2)<0$.  The
$x$-dependence of $F(x,\underline{k}^2)$ with fixed value of
$\underline{k}^2$ using Eq. (5.2) is illustrated by the solid curves
in Fig. 14. Surprisedly, the results show that
$F(x,\underline{k}^2)$ suddenly drops near a critical value $x_c\sim
1.3\times 10^{-6}$. For comparison, we calculate the BFKL equation
(2.35) and BK-like equation (4.18) with the same input. The
corresponding solutions are presented by the pointed and dashed
curves.

  \vskip -3.0 truecm \hbox{
\centerline{\epsfig{file=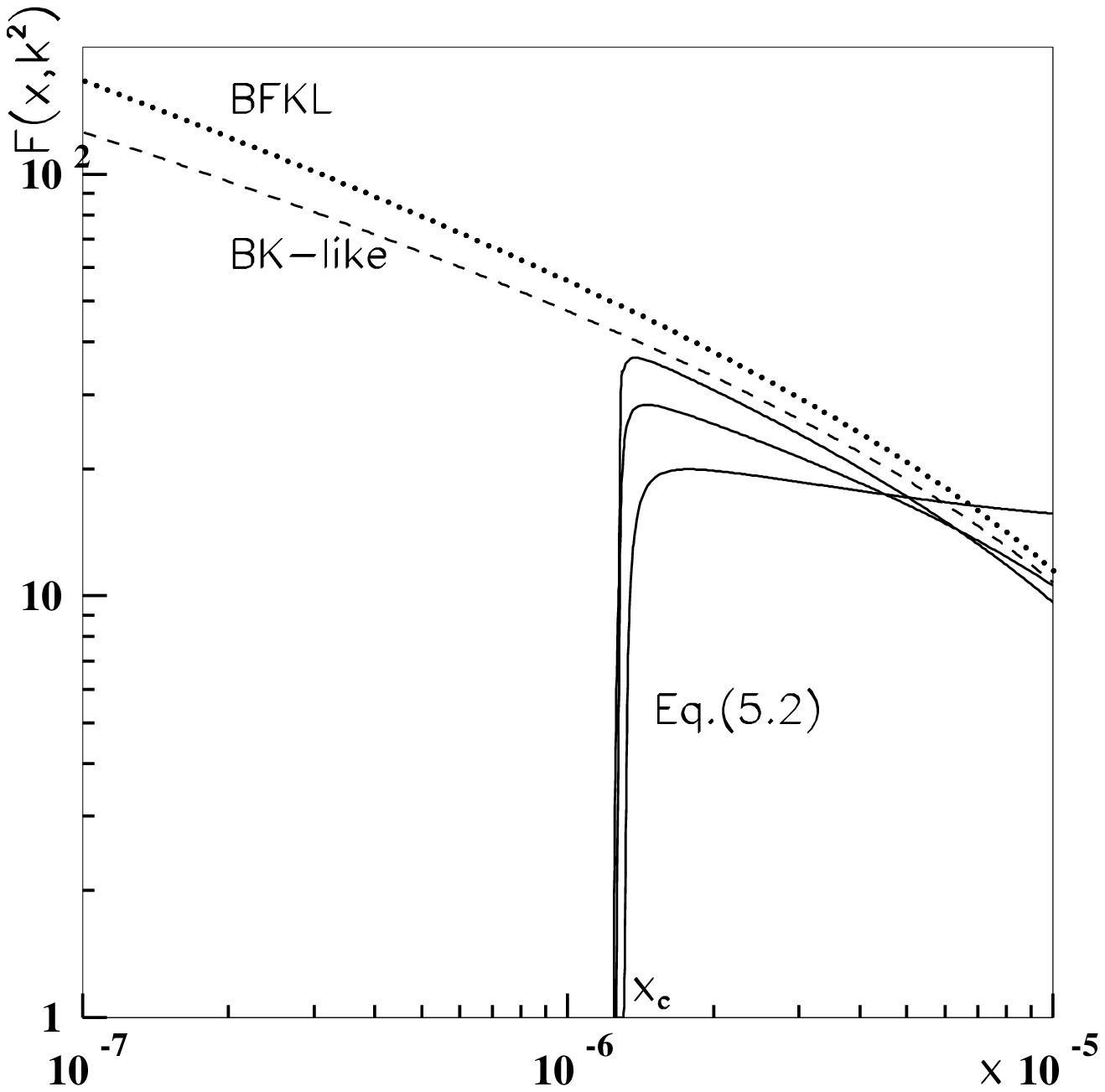,width=12.0cm,clip=}}} \noindent
\vskip -4.5 truecm

Fig. 14 $x$-dependence of the unintegrated gluon distribution in
Eq. (5.2) with the GBW input (5.3)+(5.4); the solid curves: (from
top) $\underline{k}^2=$50, 10 and 2 $GeV^2$. The results show that
the evolution of $F(x,\underline{k}^2)$ is blocked in Eq. (5.2)
near $x_c\sim 1.3\times 10^{-6}$. The dotted and dashed curves are
the corresponding solutions of the BFKL equation (2.35) and
BK-like equation (6.7) with $\underline{k}^2=50GeV^2$.

\vskip -4.0 truecm \hbox{
\centerline{\epsfig{file=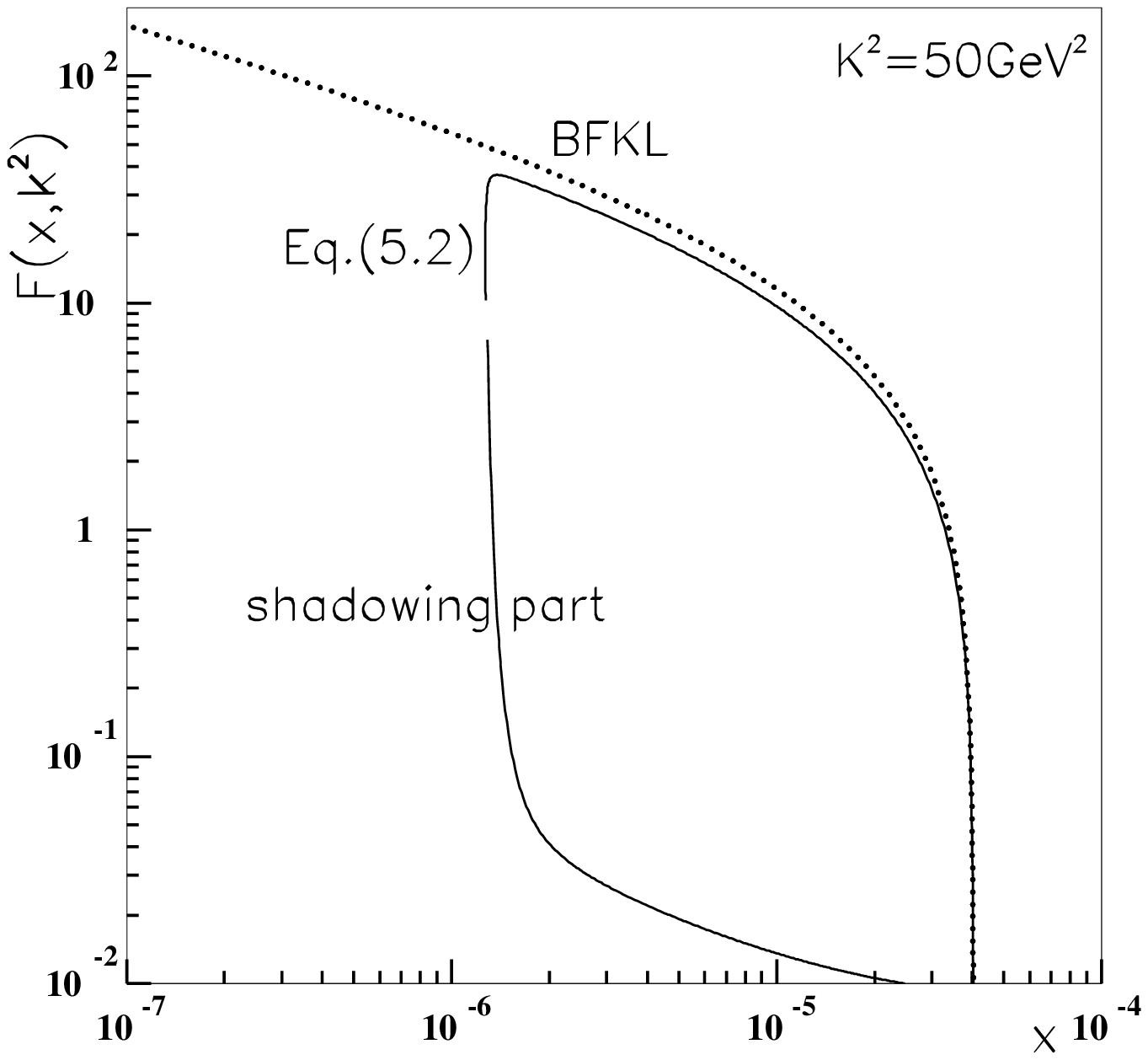,width=12.0cm,clip=}}} \noindent
\vskip -4.5 truecm

\noindent Fig. 15 The same solutions in Fig. 14, but where the
contributions from the nonlinear shadowing part in Eq. (5.2) are
separately indicated. The results show that the shadowing effect
increases suddenly near $x_c$.

 \vskip 1.0 truecm

    We plot the contributions from the nonlinear shadowing
terms of Eq. (5.2) separately in Fig. 15 and compare them with the
results of the BFKL equation. We find that the shadowing effect
increases suddenly in Eq. (5.2) near $x_c$. That is, the QCD
evolution is blocked by an anomalous shadowing effect in Eq. (5.2).

    We use the $\underline{k}^2$-dependence of $F(x,
\underline{k}^2)$ in Fig. 16 to expose the origin of the QCD
evolution block. The curves show the aperiodic oscillation and even
a dramatic change of $F(x, \underline{k}^2)$ when $x$ goes to $x_c$
near the saturation scale $\underline{k}^2\sim Q^2_s$.

\vskip -4.0 truecm \hbox{
\centerline{\epsfig{file=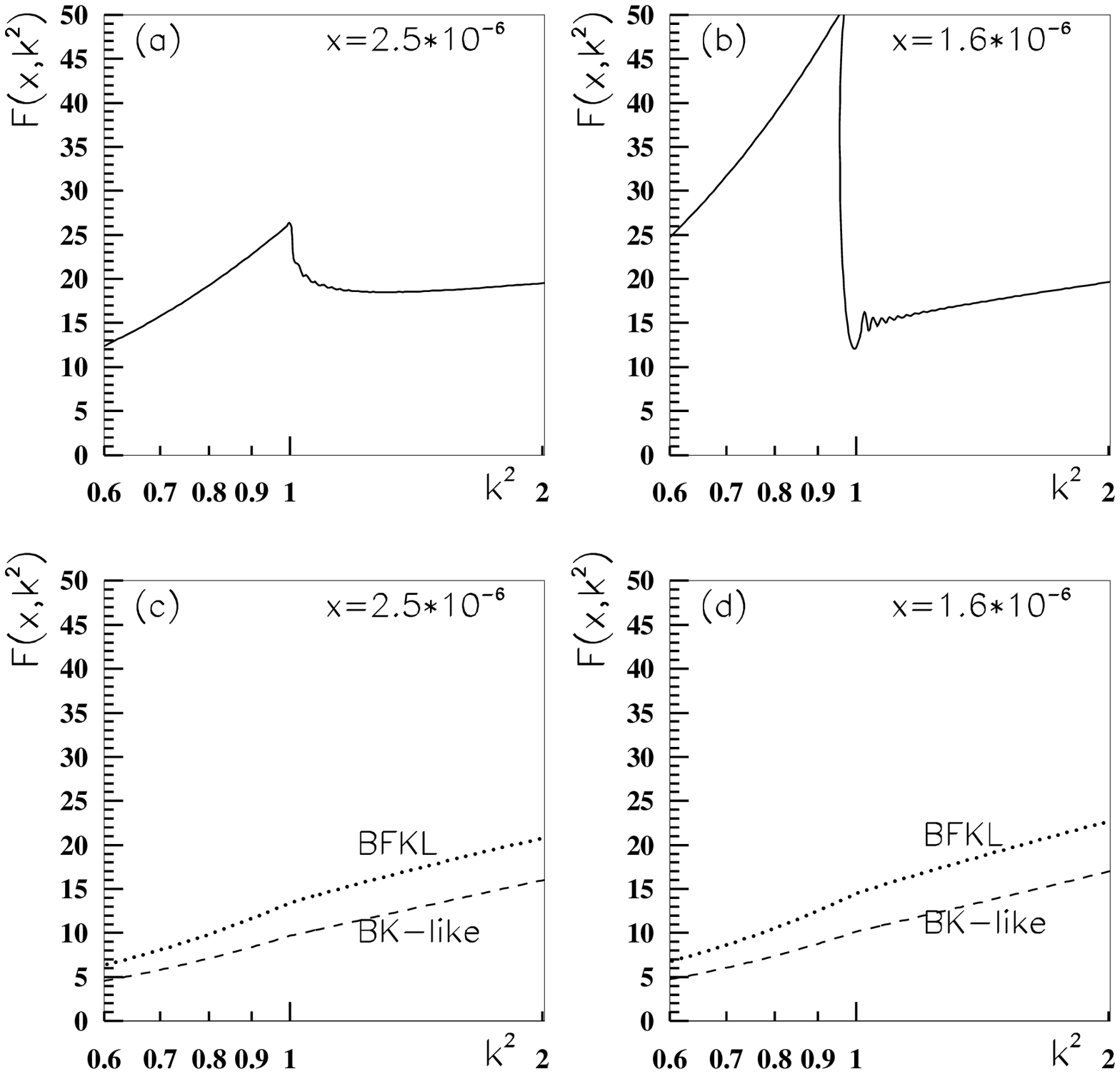,width=12.0cm,clip=}}} \noindent
\vskip -0.5 truecm

Fig. 16 (a) (b) $\underline{k}^2$-dependence of the unintegrated
gluon distribution for two different values of $x$ (solid curves);
The dotted and dashed curves are the corresponding solutions of
the BFKL and BK-like equations; (c) and (d) are parts of (a) and
(b), respectively.

 \vskip 1.0 truecm

The sudden change of a solution is an interesting phenomenon in nonlinear
evolution system, in particular, this behavior perhaps relates to
chaos. An important character of chaos is that the solution is
sensitively relevant to the initial conditions. For this sake, we
study the solutions of different input conditions. We compute a
similar solution as Fig. 16 but the starting point is moved a little
from $x_0=0.4\times 10^{-4}$ to $x_0=0.35\times 10^{-4}$. The
results in Fig. 17 show that the oscillation structure of
$F(x,\underline{k}^2)\sim\underline{k}^2$ is sensitive to the
starting point of the evolution, although the global behaviors of
the curves are similar.

    We change the input distribution to $1.01\times$Eq. (5.4) and compare these results in Fig. 18.
One can find the obvious difference in the oscillation structure.

\vskip -4.0 truecm \hbox{
\centerline{\epsfig{file=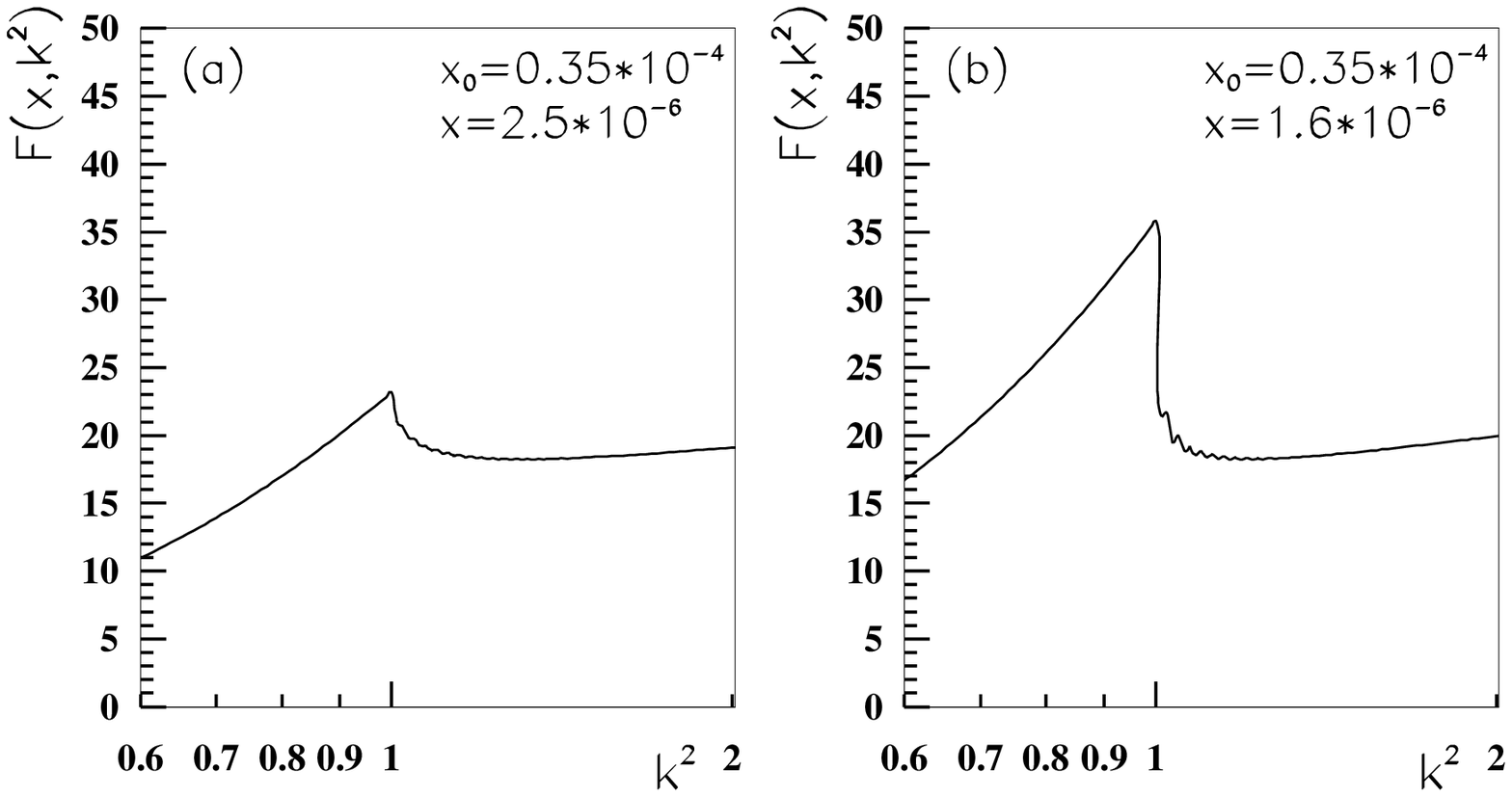,width=12.0cm,clip=}}} \noindent
\vskip -3.0 truecm

Fig.17: Comparing with Fig. 16 but evolving from $x_0=0.35\times
10^{-7}$. The results show that the oscillation structure is
sensitive to the starting point $x_0$ of the evolution.

 \vskip 1.0 truecm

\vskip -3.0 truecm
 \vskip 0 truecm \hbox{
\centerline{\epsfig{file=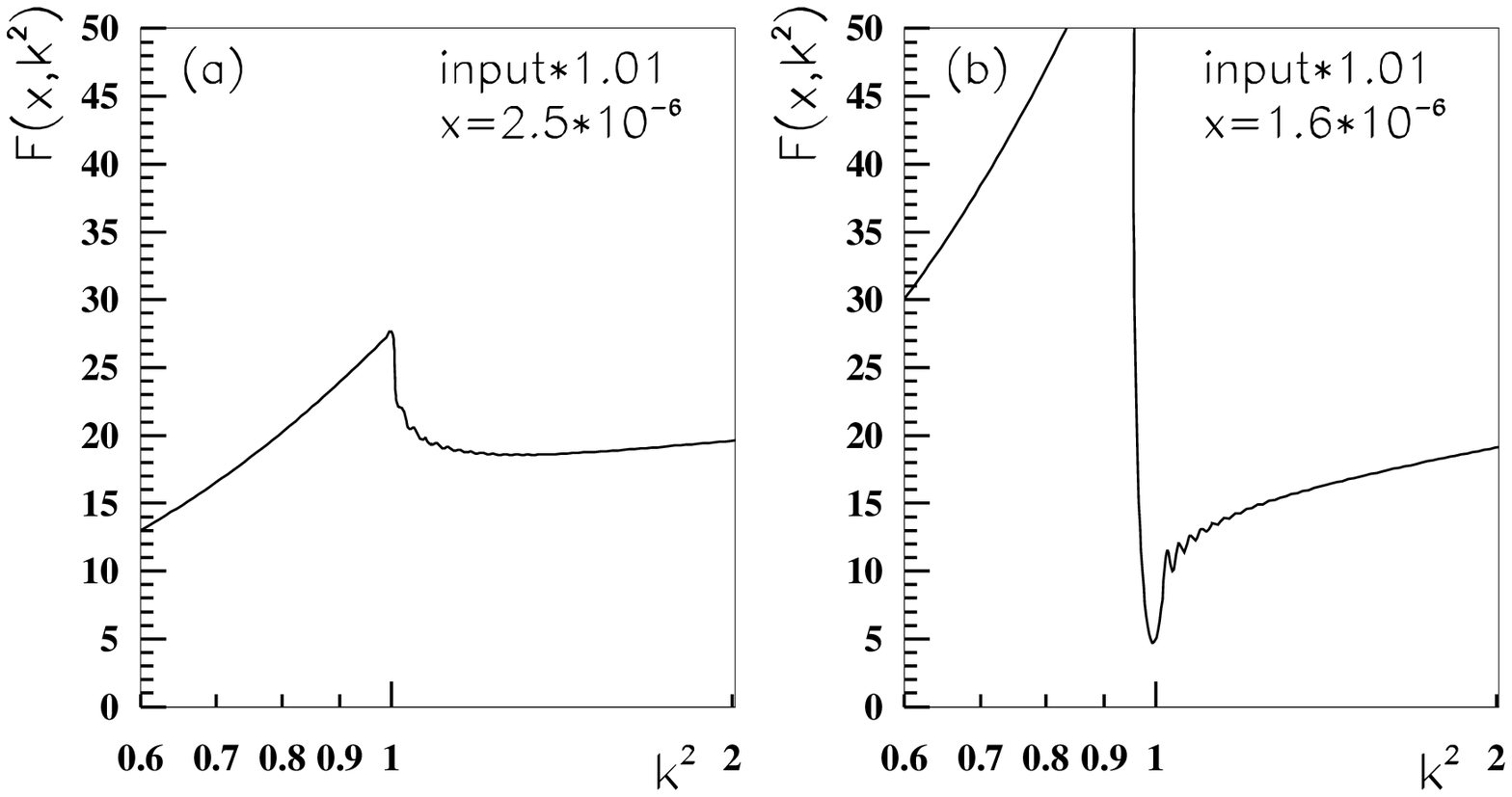,width=12.0cm,clip=}}} \noindent
\vskip -3.0 truecm

Fig.18: Comparing with Fig. 16 but using $1.01\times$ input. The
results show that the oscillation structure is sensitive to the
input distribution.

 \vskip 1.0 truecm

\vskip -4.0 truecm \hbox{
\centerline{\epsfig{file=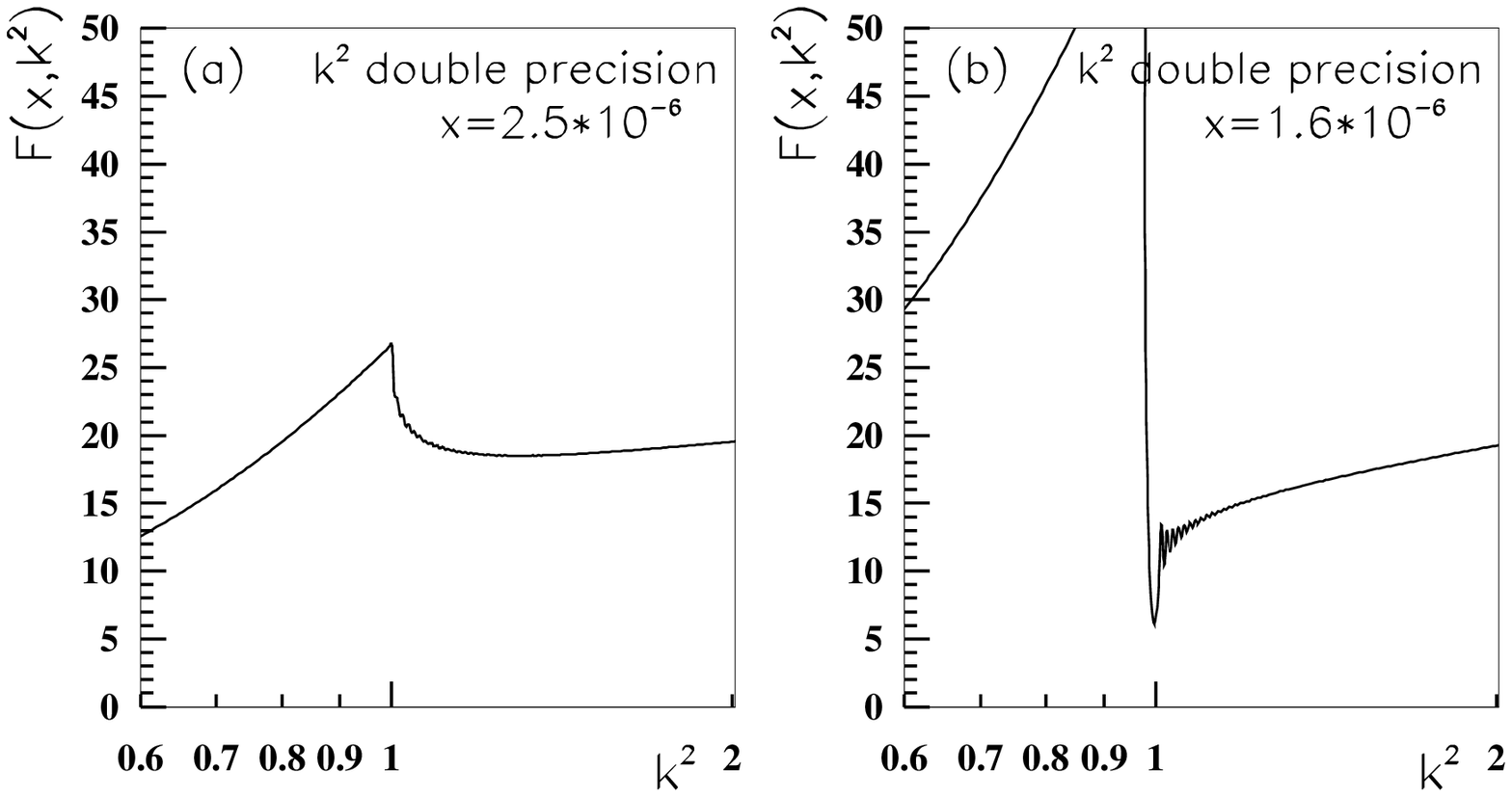,width=12.0cm,clip=}}} \noindent
\vskip -3.0 truecm

Fig.19: Comparing with Fig. 16 but using the double precisions.
The results show that the oscillations are aggravated with
increasing precision.

 \vskip 1.0 truecm

 The improvement of the precision in the computation may aggravate the chaotic
oscillations since the increasing samples perturb the distributions
at every step in the evolution. In contrast, if the above mentioned
oscillations are arisen from the calculation errors, such
oscillations will disappear with the increasing precision. In Fig.
19 we present the curve with a same input as in Fig. 16 but with
double calculating precision. One can find that the oscillations are
aggravated with increasing precision.

    The above aperiodic oscillation is sensitive to the initial
conditions. Especially, the oscillation will be enhanced with the
increase of the numerical calculation precision. These features are
universally observed in many chaos phenomena.

    A standard criterion of chaos is that the system has the positive
Lyapunov exponents, which indicates a strong sensitivity to small
changes in the initial conditions [6]. We regard $y=\ln 1/x$ as
`time' and calculate the Lyapunov exponents
$\lambda(\underline{k}^2)$ in a finite region, where the
distribution oscillation is obvious. We divide equally the above
mentioned $y$-region into n parts with $y_1,y_2.., y_{n+1}$ and
$\tau=(y_{n+1}-y_1)/n$. Assuming that the distribution evolves to
$y_1$ from $y_0=\ln 1/x_0$ and results $F(y_1,\underline{k})$.
Corresponding to a given value $F(y_1,\underline{k})$ at
$(y_1,\underline{k})$, we perturb it to
$F(y_1,\underline{k})+\Delta$ with $\Delta\ll 1$. Then we continue
the evolutions from $F(y_1,\underline{k})$ and
$F(y_1,\underline{k})+\Delta$ to $y_2$ from $y_1$ respectively, and
denote the resulting distributions as $F(y_2,\underline{k})$ and
$\tilde{F}(y_2,\underline{k})$. Making the difference
$\Delta_2=\vert\tilde{F}(y_2,\underline{k})-F(y_2,\underline{k})\vert$.
In the following step, we repeat the perturbation
$F(y_2,\underline{k})\rightarrow F(y_2,\underline{k})+\Delta$ and
let the next evolutions from $F(y_2,\underline{k})$ and
$F(y_2,\underline{k})+\Delta$ from $y_2$ to $y_3$ respectively and
get the results $\Delta_3=
\vert\tilde{F}(y_3,\underline{k})-F(y_3,\underline{k})\vert$......
(see Fig. 20). The Lyapunov exponents for the image from $y$ to
$F(y,\underline{k})$ are defined as

$$\lambda(\underline{k}^2)=\lim_{n\rightarrow\infty}\frac{1}{n\tau}\sum_{i=2}^{n+1}\ln\frac {\Delta_i}
{\Delta}. \eqno(5.5)$$ The Lyapunov exponents of the gluon
distribution in Eq. (5.2) with the input Eq. (5.4) are presented in
Fig. 21. For comparison, we give the Lyapunov exponents of the BFKL
and BK-like equations. The positive values of the Lyapunov exponents
clearly show that the oscillation of $F(x,
\underline{k}){\sim}\underline{k}^2$ is chaos of Eq. (5.2).
Therefore, we conclude that chaos in Eq. (5.2) blocks the QCD
evolution of the gluon distribution.

\vskip -6.0 truecm \hbox{
\centerline{\epsfig{file=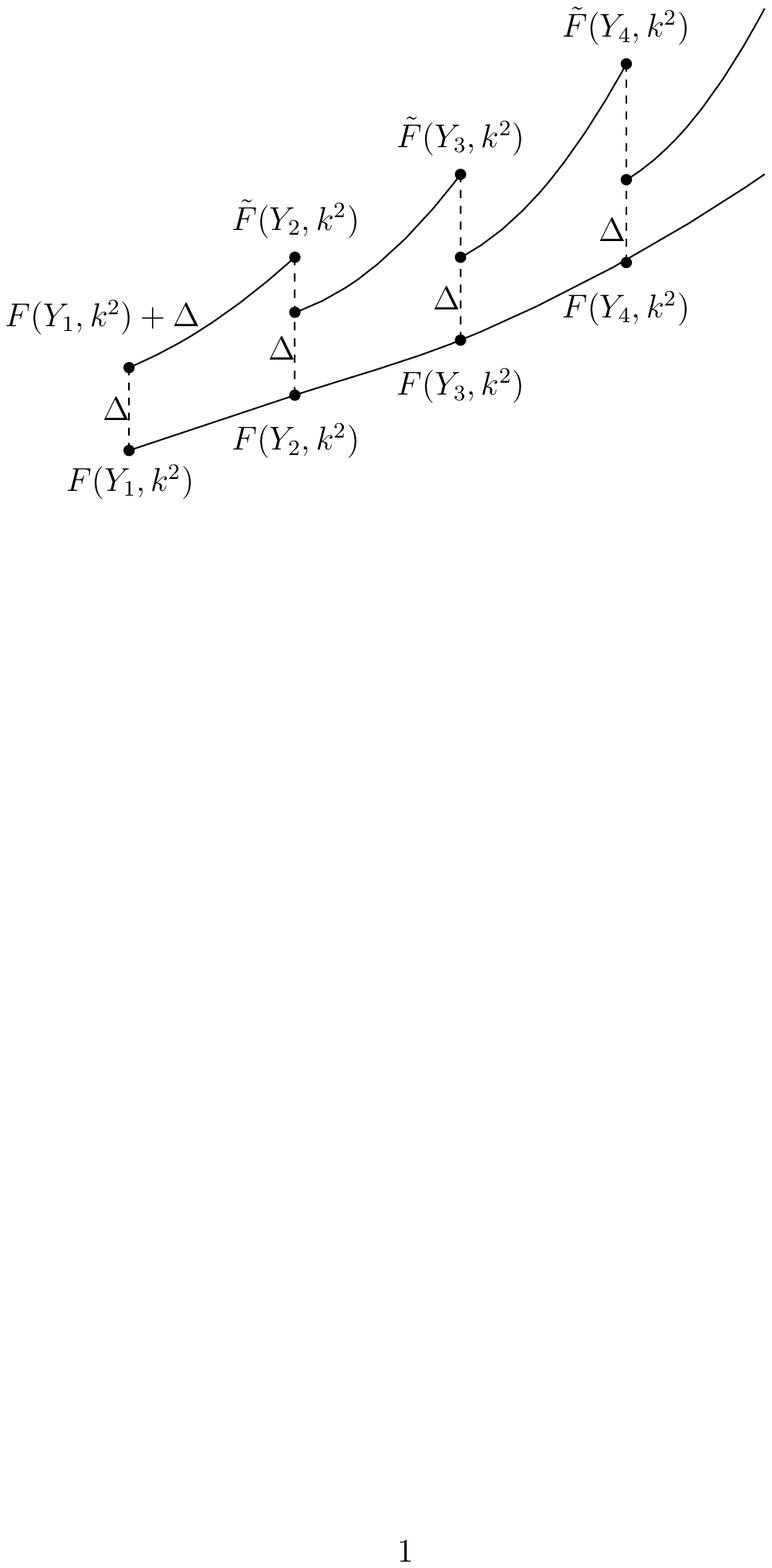,width=12.0cm,clip=}}} \noindent
\vskip -10.0 truecm

Fig.20: Schematic programs to calculate the Lyapunov exponents of
the evolution equations.

\vskip -0.0 truecm \hbox{
\centerline{\epsfig{file=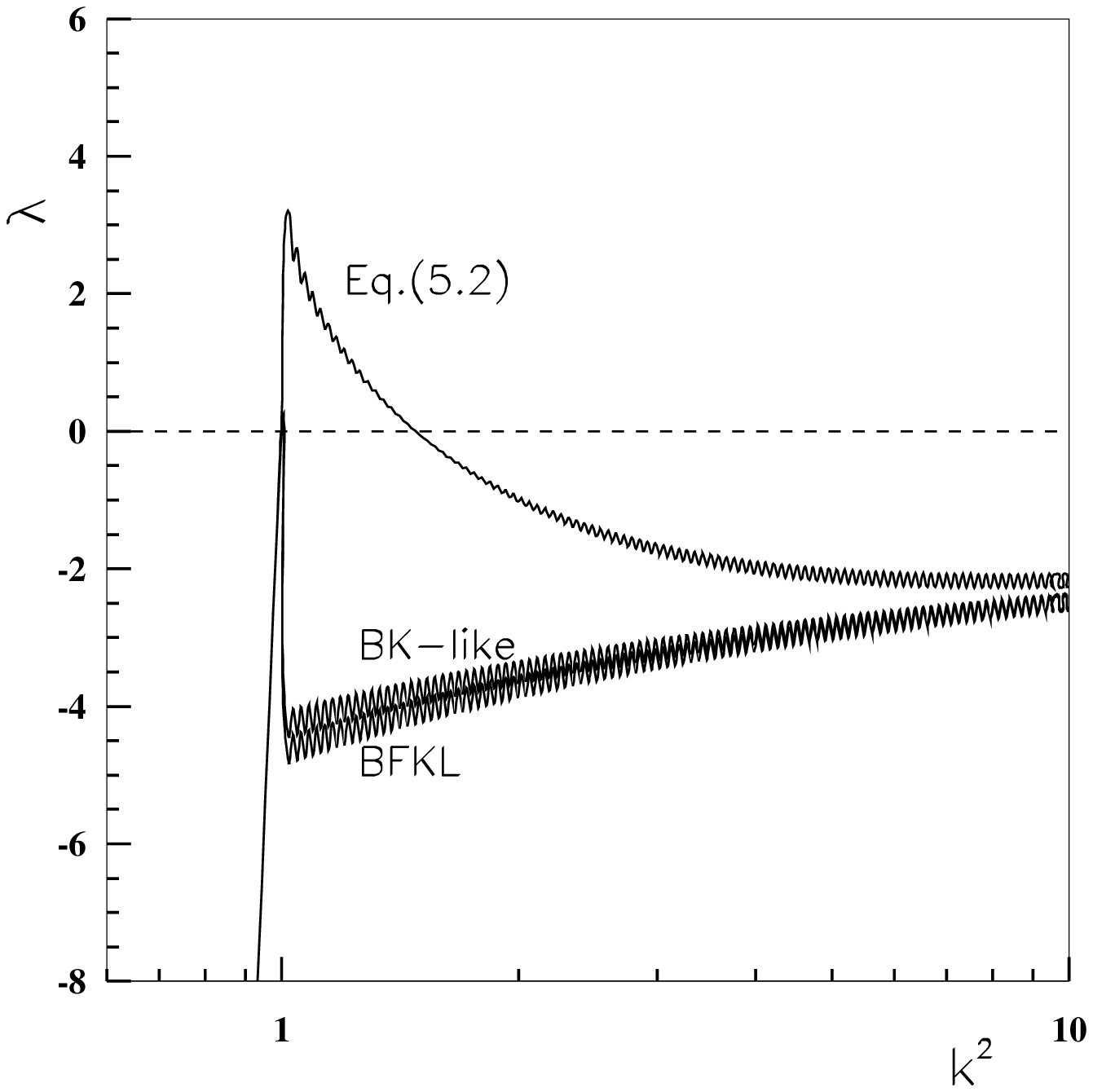,width=12.0cm,clip=}}} \noindent
\vskip -2.0 truecm

Fig.21: The positive Lyapunov exponents show that the
corresponding solution of Eq. (5.2) is chaos.

\newpage
\begin{center}
\section{Discussions}
\end{center}

    The exact value of $x_c$ depends on
the initial conditions, which have some uncertainties, however, the
fact of chaos is irrelevant to the detailed dynamics, provided an
essential change of the $\underline{k}^2$-dependence of
$F(x,\underline{k}^2)$ when the evolution transfers from
perturbative to nonperturbative ranges. For example, an alternative
saturation model [19] assumes that

$${\mathcal{F}}(x_0,\underline{k}^2)\rightarrow ~ constant, at~\underline{k}^2\leq Q^2_s. \eqno(6.1)$$
We use

$$F(x,\underline{k}^2)=C k^2/(k^2+k^2_a),~~at~x\leq x_0,~\underline{k}^2\leq Q^2_s,\eqno(6.2)$$
with $k^2_a=1GeV^2$ to replace Eq. (5.4) (for $x=x_0$, see the
dashed curve in Fig. 13). The chaos solutions still exist in Fig. 22
where $x_c\sim 1.7\times 10^{-7}$. The reason of the chaos solution
in Eq. (5.2) is that this equation contains the following
regularized kernels

$$\left[\frac{F(x,\underline{k}'^2)}{\vert
\underline{k}'^2-\underline{k}^2\vert}-\frac{F(x,\underline{k}^2)}{\vert
\underline{k}'^2-\underline{k}^2\vert}\right]_{\underline{k'}^2\sim\underline{k}^2}\sim
\frac{d}{d\underline{k}'^2}\left[
F(x,\underline{k}'^2)\right]_{\underline{k'}^2\sim\underline{k}^2},
\eqno(6.3)$$ in the linear terms and

$$\left[\frac{\underline{k}^2F^2(x,\underline{k}'^2)}{\underline{k}'^2\vert
\underline{k}'^2-\underline{k}^2\vert}-\frac{\underline{k}'^2F^2(x,\underline{k}^2)}{\underline{k}'^2\vert
\underline{k}'^2-\underline{k}^2\vert}\right]_{\underline{k'}^2\sim\underline{k}^2}\sim
\frac{d}{d\underline{k}'^2}\left[\frac{\underline{k}^2}
{\underline{k}'^2}
F^2(x,\underline{k}'^2)\right]_{\underline{k'}^2\sim\underline{k}^2},
\eqno(6.4)$$ in the nonlinear terms. The derivation of
$F(x,\underline{k}^2)$ with respect to $\underline{k}^2$ adds a
perturbation on the smooth curve $F(x,\underline{k}^2)$ once
$\underline{k}$ crosses over $Q_s$. Thus, we have a serious of
independent perturbations in a narrow $\underline{k}^2$ domain along
$x$ ($x<x_0$). In the linear BFKL equation, these perturbations are
independent and their effects are negligibly small. The solutions
keep the smooth curves both on the $x$- and $\underline{k}^2$-spaces
as shown in Figs. 16cd. However, the nonlinear Eq. (5.2) may occur
the coupling among these random perturbations and forms chaos near
$Q^2_s$. Although we don't yet know this detail, the positive
Lyaponov exponents of Eq. (5.2) in Fig. 21 strongly support our
suggestion. The distribution $F(x,\underline{k}^2)$ is an evolution
result from $F(x-\Delta,\{\underline{k}^2\})$, where
$\{\underline{k}^2\}$ overlaps a whole kinematic range including
$\underline{k}^2=Q^2_s$. Once chaos is produced near $x\sim x_c$ and
$\underline{k}^2\sim Q^2_s$, the fast oscillations of the gluon
density arise a huge shadowing due to Eq. (6.4), and the evolution
of the distribution $F(x,\underline{k}^2)$ is suddenly blocked near
$x_c$. The normal shadowing in the GLR-MQ-ZRS and BK equations
origins from a large value of the gluon distribution, while the big
shadowing in Eq. (5.2) is arisen by the rapid oscillations of the
chaos solution. We call this new shadowing as the blocking effect.

\vskip -3.0 truecm \hbox{
\centerline{\epsfig{file=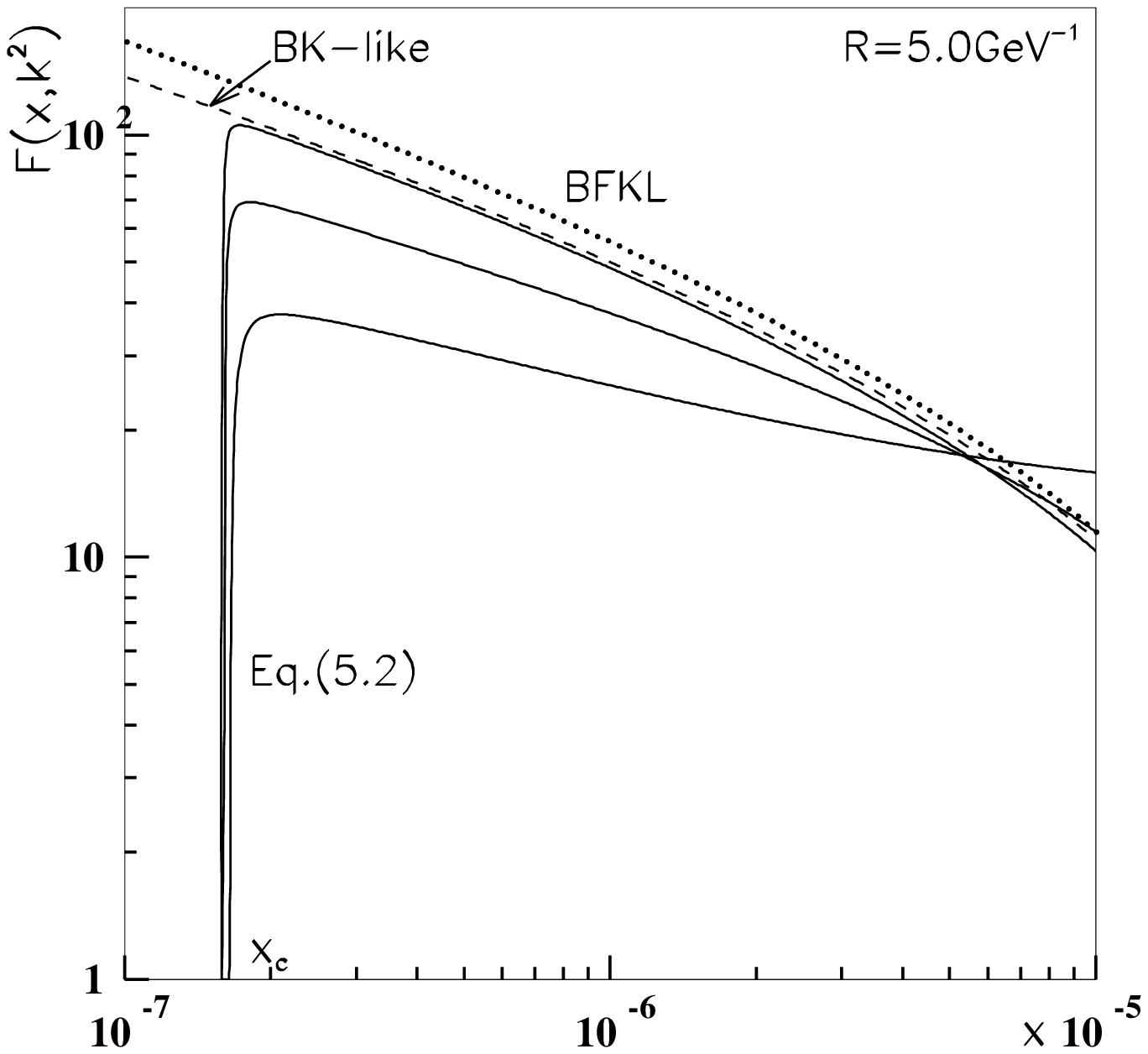,width=12.0cm,clip=}}} \noindent
\vskip -4.5 truecm

 Fig.22: Similar to Fig. 14 but the input (5.4) is replaced
by the saturation model (6.2).

\vskip 1.0 truecm

    We discuss qualitatively the azimuthal angle $(\phi)$-dependent case. Equation (5.2) becomes

$$-x\frac{\partial F(x,\underline{k})}{\partial x}$$
$$=\frac{3\alpha_{s}\underline{k}^2}{\pi}\int_{\underline{k}^2_0}^{\infty} \frac{d^2 \underline{k
}'}{\underline{k}'^2}\left\{\frac{F(x,\underline{k}')-F(x,\underline{k})}
{(\underline{k}'-\underline{k})^2}+......\right\}$$
$$-\frac{81}{16}\frac{\alpha_s^2}{\pi R^2_N}\int_{\underline{k}^2_0}^{\infty} \frac{d^2 \underline{k
}'}{\underline{k}'^2}\left\{\frac{\underline{k}^2F^2(x,\underline{k}')-\underline{k}'^2F^2(x,\underline{k})}
{\underline{k}'^2(\underline{k}'-\underline{k})^2}+......\right\},
\eqno(6.5)$$ where (......) are the non-singular parts.
One can find that the equation contains the similar regularized forms like
Eqs. (6.3) and (6.4):

$$\left[\frac{F(x,\underline{k}')}{(\underline{k}'-\underline{k})^2}-\frac{F(x,\underline{k})}
{(\underline{k}'-\underline{k})^2}\right]_{\underline{k'}\sim\underline{k}}\subseteq
\frac{\delta}{\delta\underline{k}'}\left[
F(x,\underline{k}')\right]_{\underline{k'}\sim\underline{k}},
\eqno(6.6)$$ in the linear terms and

$$\left[\frac{\underline{k}^2F^2(x,\underline{k}')}{\underline{k}'^2(
\underline{k}'-\underline{k})^2}-\frac{\underline{k}'^2F^2(x,\underline{k})}{\underline{k}'^2(
\underline{k}'-\underline{k})^2}\right]_{\underline{k'}\sim\underline{k}}\subseteq
\frac{\delta}{\delta\underline{k}'}\left[\frac{\underline{k}^2}
{\underline{k}'^2}
F^2(x,\underline{k}')\right]_{\underline{k'}\sim\underline{k}},
\eqno(6.7)$$ in the nonlinear terms. As we have emphasize that both the chaos and the blocking effect origin from such kind of regularized
kernels. Therefore, we consider that our results are still hold for the azimuthal angel-dependent solutions.

    The equation (3.46) is based on the leading QCD
corrections, where the higher order corrections are neglected. An
important questions is: will the chaos effects in the new evolution
equation disappear after considering higher order corrections? We have
known that chaos in the MD-BFKL equation origins from the
singularity of the nonlinear evolution kernel. From the experiences
in the study of the BFKL equation, higher order QCD corrections can
not remove the singularities at the lower order approximation [20]. In
particular, the virtual cut diagrams always exit in any higher
order corrections to the BFKL equation. The regularization similar
to Eq. (6.4) is necessary. Besides, the chaotic behavior cannot be destroyed
by arbitrarily small perturbations of the system parameters.
Therefore, we expect that chaos still
exists in Eq. (3.46) even considering the higher order
corrections.

    The solution $F(x,\underline{k}^2)$ of Eq. (5.2)
becomes zero can not be simply explained as the gluon disappearance
at $x\leq x_c$. Although the three gluons vertex stops working at
$x<x_c$, the gluons still can evolve similar to the Abliean photons
in a thin parton system. In a quark confinement mechanism, the
dual-superconductor picture was suggested by Refs. [21-23], where an
assumption of Abelian dominance seems to be significant to
confinement. The Abelian dominance means that only the diagonal
gluon component in the confinement mechanism. The distributions of the non-Abelian gluons
collapse at $x<x_c$, the contributions of the Abelian gluons appear.
One can image that the Abelian gluons dominate the soft gluons.
Thus, chaos in Eq. (3.46) provides a dynamical mechanism for
separating the Abelian gluons.

    The blocking effect in the QCD evolution will suppress the new
particle events in an ultra high energy hadron collision. Although
we have not exactly predicted the energy scale $x_c$ which
corresponding to the blocking effect, the chaos solutions in Eq.
(3.46) should arise our attention when considering the future large
hadron collider. In particular, the nonlinear coefficients in the
evolution equation will be enhanced by a factor
$[1+0.21(A^{1/3}-1)]$ in the nuclear target since the correlations
of gluons among different bound nucleons [24], this will increase
the value of $x_c$ into the observable range of the projected Large
Hadron Electron Collider (LHeC) [25], Very Large Hadron Collider
(100TeV VLHC) [26] and the upgrade (CepC, CppC) in a circular $e^+e^-$ collider (SppC) [27].
Figure 23 presents the nuclear A-dependence of
$x_c$. We will detail them elsewhere.

\vskip -3.0 truecm \hbox{
\centerline{\epsfig{file=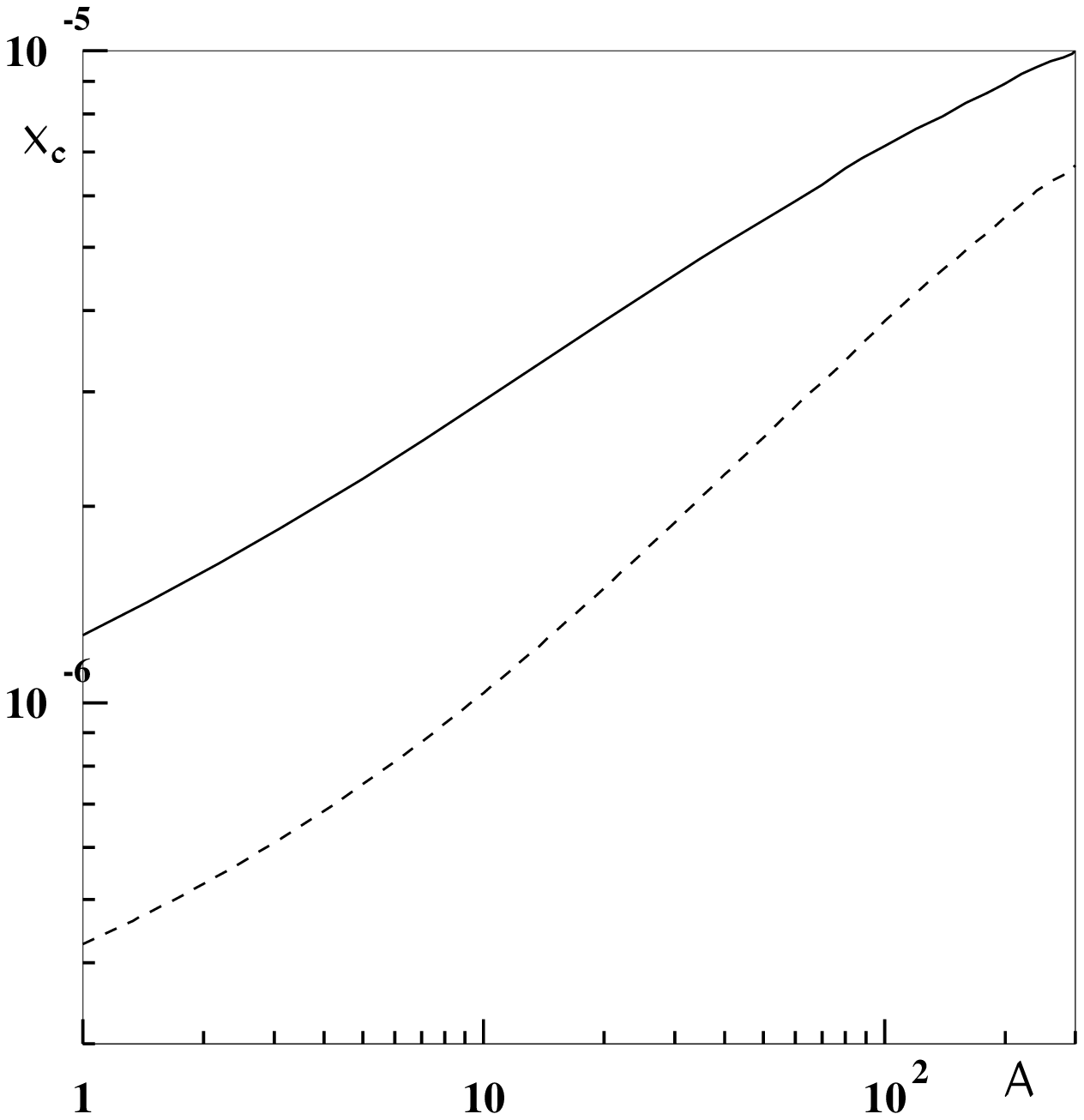,width=12.0cm,clip=}}} \noindent
\vskip -4.5 truecm

  Fig.23: Nuclear A-dependence of $x_c$ using Eq. (5.2) but
added the nuclear factor $[1+0.21(A^{1/3}-1)]$ on the nonlinear
terms. Solid curve: using input (5.4); Dashed curve: using input
(6.2).

\vskip 1.0 truecm

In summary, we derive a new evolution equation in a
unified partonic framework, where the TOPT cutting rule is used to
sum the contributions from the relating cut diagrams. This new
evolution equation sums both the leading $\ln(1/x)$ gluon splitting
and recombination contributions. We indicate that the new evolution
equation is a natural expansion of the well-known DGLAP, BFKL,
GLR-MQ-ZRS and BK equations.

    We find that the new evolution equation has the chaos solution with positive
Lyaponov exponents in the perturbative range. We indicate that chaos
in this evolution equation origins from a serious of perturbations
when the evolution crosses over the saturation scale. The fast
aperiodic oscillation of gluon distribution with $\underline{k}$ in
chaos leads to a big shadowing in the new evolution equation. This
new kind of shadowing effect may block the QCD evolution vis three
gluon vertex at small $x$. We point out that the above mentioned
chaos and blocking effects relating to the singular structure of the nonlinear
evolution kernel in the evolution equation, where the regularization with the virtual cut diagrams
is necessary.

    Although the position of
chaos is undetermined due to the value of $x_c$ sensitively
dependent on the input conditions, the existence of chaos in the QCD
evolution equation may change our expectation to the future large
hadron collider plans.

\end{document}